%% file: main.tex
\PassOptionsToPackage{table,xcdraw}{xcolor}
\documentclass[conference]{IEEEtran}
\let\labelindent\relax

\usepackage{lineno}
\usepackage{array}
\newcolumntype{C}[1]{>{\centering\arraybackslash}m{#1}}
\usepackage{CJKutf8}
\input{command}
\usepackage[normalem]{ulem}
\usepackage{multirow}
\usepackage{graphicx}
\usepackage[caption=false]{subfig}
\usepackage[textsize=tiny]{todonotes}
\usepackage{enumitem}
\usepackage{amsmath}
\usepackage{amssymb}
\usepackage[hidelinks]{hyperref}
\usepackage{cleveref}
\usepackage[justification=justified]{caption}
\usepackage{adjustbox}

\hypersetup{
  colorlinks   = true,    
  urlcolor     = red,    
  linkcolor    = red,    
  citecolor    = red      
}

\newcommand{\hl}[1]{\textcolor{black}{#1}}

\ifCLASSINFOpdf
\else
\fi
\begin{document}
%


\title{Detecting Voice Cloning Attacks via Timbre Watermarking}

\author{ \normalsize
	{Chang Liu}\mkustc,
	{Jie Zhang}\mkntu \mkletter{},
	{Tianwei Zhang}\mkntu,
    {Xi Yang}\mkustc,
	{Weiming Zhang}\mkustc \mkletter{}, 
	{and Nenghai Yu}\mkustc \\
 \normalsize
	\mkustc {University of Science and Technology of China}\\
 \normalsize
	\mkntu {Nanyang Technological University} \\
 \normalsize
\textit{\mkletter{}Corresponding Authors}\\
\textit{  \{hichangliu@mail., yx9726@mail., zhangwm@, ynh@\}ustc.edu.cn,  \{jie\_zhang, tianwei.zhang\}@ntu.edu.sg}
}

\IEEEoverridecommandlockouts
\makeatletter\def\@IEEEpubidpullup{6.5\baselineskip}\makeatother
\IEEEpubid{\parbox{\columnwidth}{
    Network and Distributed System Security (NDSS) Symposium 2024\\
    26 February - 1 March 2024, San Diego, CA, USA\\
    ISBN 1-891562-93-2\\
    https://dx.doi.org/10.14722/ndss.2024.24200\\
    www.ndss-symposium.org
}
\hspace{\columnsep}\makebox[\columnwidth]{}}


\maketitle

\input{0_abstract}
\input{1_introduction}
\input{2_background}
\input{3_preliminary}
\input{4_method}

\input{5_experiments}
\input{6_conclusion}

\vspace{3mm}
\small

\bibliographystyle{IEEEtran}
\bibliography{sample-base}

\input{7_appendix}

\end{document}

%% file: command.tex
\usepackage{color}
\usepackage{pifont}

\newcommand{\circled}[1]{\textcircled{\raisebox{-0.9pt}{#1}}}

\definecolor{green}{rgb}{0, 0.5, 0}
\definecolor{orange}{rgb}{0.8, 0.6, 0.2}
\definecolor{orange2}{rgb}{1.0, 0.6, 0.2}
\definecolor{red}{rgb}{1.0, 0.0, 0.0}
\definecolor{teal}{rgb}{0.0, 0.4, 0.4}
\definecolor{purple}{rgb}{0.65,0,0.65}
\definecolor{saffron}{rgb}{0.95,0.75,0.2}
\definecolor{turquoise}{rgb}{0.0,0.5,0.5}
\definecolor{black}{rgb}{0.0, 0.0, 0.0}
\definecolor{gray}{rgb}{0.5, 0.5, 0.5}

\newcommand{\etal}{\textit{et al}. }

\newcommand{\ie}{\textit{i}.\textit{e}.}
\newcommand{\eg}{\textit{e}.\textit{g}.}

\newcommand{\Tref}[1]{\textcolor{red}{Table~\ref{#1}}}
\newcommand{\Eref}[1]{\textcolor{red}{Eq.~(\ref{#1})}}
\newcommand{\Fref}[1]{\textcolor{red}{Fig.~\ref{#1}}}
\newcommand{\Sref}[1]{\textcolor{red}{Sec.~\ref{#1}}}
\newcommand{\ASref}[1]{\textcolor{red}{Appx.~\ref{#1}}}

\newcommand{\para}[1]{{\vspace{1.5pt} \bf \noindent #1 \hspace{4pt}}}

\newcommand{\mkustc}[0]{{{$^1$}}}
\newcommand{\mkntu}[0]{{{$^2$}}}
\newcommand{\mkletter}[0]{{{\normalsize \textsuperscript{\dag}}}}

%% file: 0_abstract.tex
\begin{abstract}

Nowadays, it is common to release audio content to the public, for social sharing or commercial purposes. 
However, with the rise of voice cloning technology, attackers have the potential to easily impersonate a specific person by utilizing his publicly released audio without any permission.
Therefore, it becomes significant to detect any potential misuse of the released audio content and protect its timbre from being impersonated.

To this end, we introduce a novel concept, ``Timbre Watermarking'', which embeds watermark information into the target individual's speech, eventually defeating the voice cloning attacks. 
However, there are two challenges: 1) \textit{robustness}: the attacker can remove the watermark with common speech preprocessing before launching voice cloning attacks; 2) \textit{generalization}: there are a variety of voice cloning approaches for the attacker to choose, making it hard to build a general defense against all of them. 

To address these challenges, we design an end-to-end voice cloning-resistant detection framework. 
The core idea of our solution is to embed the watermark into the frequency domain, 
which is inherently robust against common data processing methods. A repeated embedding strategy is adopted to further enhance the robustness. To acquire generalization across different voice cloning attacks, we modulate their shared process and integrate it into our framework as a distortion layer.
Experiments demonstrate that the proposed timbre watermarking can defend against different voice cloning attacks, exhibit strong resistance against various adaptive attacks (\hl{\eg, reconstruction-based removal attacks, watermark overwriting attacks}), and achieve practicality in real-world services such as PaddleSpeech, Voice-Cloning-App, and so-vits-svc. In addition, ablation studies are also conducted to verify the effectiveness of our design. Some audio samples are available at \url{https://timbrewatermarking.github.io/samples}.

\end{abstract}

%% file: 1_introduction.tex
\section{Introduction}

\begin{quote}
\textit{``The voice is an instrument that you can learn to play and use in a way that is uniquely yours.'' }
\begin{flushright}
\textbf{-- Kristin Linklater}
\end{flushright}
\end{quote}

We are already in the era of the Ear Economy, where audio content has become increasingly popular in today's digital landscape. Many individuals enjoy sharing their voice artworks (\eg, music recordings, audio books, soundtracks) on public platforms, such as Spotify \cite{spotify}, Audible \cite{audible}, Himalaya \cite{himalaya}, SoundCloud \cite{soundcloud}, etc. Those services have significantly changed the way people consume content for entertainment and gaining knowledge. 

\begin{figure}[t] 		
	\centering	
	\includegraphics[width=1\linewidth]{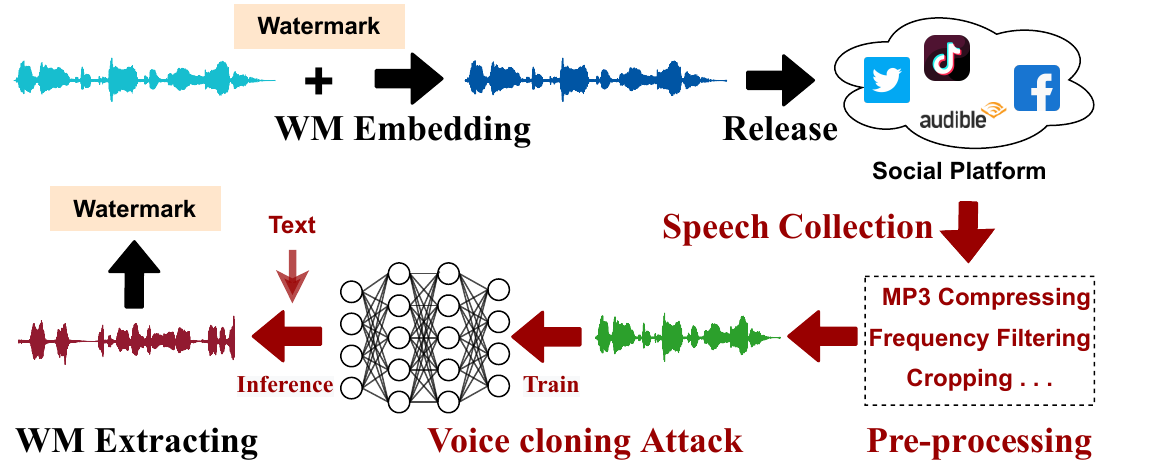}
	\caption{The process of embedding and extracting ``Timbre Watermarking'' for detecting voice cloning attacks.}
	\vspace{-15pt}
	\label{fig:f1}
\end{figure}

Unfortunately, the rise of the voice cloning technology \cite{kaneko2019cyclegan, qian2019autovc,wang2017tacotron, ren2020fastspeech} exposes new security challenges to the audio content sharing services. 
Recent advances in voice cloning leverage deep learning to accurately synthesize human voices indistinguishable from the target ones \cite{arik2017deep,oord2016wavenet}. This technology has been widely applied in different scenarios: in the entertainment industry, it can be used to create synthetic voices for characters in movies and video games; in healthcare, it can generate voices for patients who have lost their ability to speak due to illness or injury.
However, an adversary can also exploit this technology to generate voices of other persons' timbre (\ie, patterns, intonations, and other characteristics of human speech) for illegal or unauthorized purposes. This is dubbed \textit{voice cloning attack}, which can cause various severe consequences, such as financial losses, reputation damage, and copyright infringement. For instance, a clip on YouTube shows that Biden announced a plan of launching a unclear attack against Russia, which was created by voice cloning and could cause public panic \cite{fortune}.

Therefore, unauthorized synthesis of valuable timbre shall be not acceptable, and
it is crucial to establish safeguards against the potential misuse of voice cloning technology. Several attempts have been made to tackle this issue. However, they all suffer from some limitations.  First, passive detection methods \cite{ahmed2020void,gao2010audio,lieto2019hello,wang2020deepsonar} have been developed to identify whether a voice is produced using voice cloning. However, advanced voice cloning techniques make these methods obsolete, and create a perpetual arms race between voice synthesis and detection techniques \cite{wenger2021hello}. 
Second, another line of defense strategies are introduced to proactively prevent voice cloning. Specifically, Huang \etal \cite{huang2021defending} proposed to add adversarial noise to the target voice, causing the attacker to create synthetic voices with totally different timbre from the target one. However, this solution has two practical issues: (1) it needs to add significant amounts of noise to achieve the protection, which can obscure the original voice and affect its \hl{fidelity}; (2) it needs domain knowledge about the specific voice cloning approach adopted by the attacker, and cannot be generalized to other attack approaches.

To address the limitations of existing solutions, motivated by the recent deep model watermarking technology \cite{zhang2020model,zhang2021deep},
we introduce a novel concept called ``\textit{Timbre Watermarking}'' to protect the released timbre against voice cloning. As shown in \Fref{fig:f1}, we proactively embed watermark information (\eg, ownership) into the target voice before releasing it to the public. To ensure its normal usage, the watermarked voice is indistinguishable from the original one and users will not notice its existence. An attacker may collect such watermarked voice without permission, and then apply a voice cloning method to generate synthetic voice with the same timbre but different content or semantic information. By extracting the watermark information from the synthetic voice, we are able to detect the fake speech and track its original timber reliably. 

\hl{Current audio watermarking methods have demonstrated success in preserving fidelity while guaranteeing normal robustness (\eg,  against desynchronization/recapturing attacks \cite{zhao2022ssvs, liu2018patchwork}). However, they cannot be applied for timbre watermarking, due to their incapability of handling voice cloning attacks (explained in \Sref{sec:audio-wm-background}). }
Here, we conclude two challenges in realizing timbre watermarking: (1) \textit{Generalization}. We do not know how the attacker would launch the voice cloning attacks, including data collection, generative model, cloning strategies, etc. It is difficult to embed a timbre watermark that can resist all types of possible voice cloning strategies. (2) \textit{Robustness}. The attacker may attempt to remove the potential watermarks by preprocessing the voice. For instance, he can crop the voice to randomize the speech length and complicate the resynchronization for watermark extraction; he can also compress the audio signals to damage the hidden watermark. Although a number of prior works have proposed methods to watermark audio works \cite{su2018snr,liu2018patchwork,zhao2021desynchronization, liu2022dear, pavlovic2022robust}, 
as they are not targeting the voice cloning attacks, they are not robust and can be easily removed by the attacker. In \Sref{sec:appendix-comparison} we experimentally validate the limitations of these existing watermark solutions: the indeterminate length and scale of the synthesized speech will totally fail their resynchronization process before extracting, which usually depends on the synchronization codes, shifting mechanisms, or time scaling-invariant features.

In this paper, we propose a novel end-to-end timbre watermarking framework to defeat voice cloning attacks. \Fref{fig:framework} shows the overview of our methodology. Specifically, (1) to enhance the robustness, we propose to embed the watermark information into the transform domain. We adopt the Short-Time Fourier Transform (STFT) scheme to transfer the audio wave and embed the watermark into its frequency domain (vertical direction), which is inherently robust against voice processing operations along the time domain (horizontal direction). To further strengthen the watermark robustness, we repeat the embedding strategy along the horizontal direction, eliminating the dependency on the time domain again. Symmetrically, during the extraction stage, we average the extracted watermark along the horizontal direction. (2) To enhance the generalization, we investigate existing popular voice cloning strategies and find some common-used processing operations, such as scale modification, normalization, phase information discarding, and waveform reconstruction. Then, we insert these processes as a distortion layer between the watermark embedding and extraction modules and train them together in an end-to-end way. This distortion layer provides other modules the awareness of different processing operations when verifying the watermarks. After training, we discard the distortion layer and leverage the other modules for watermark embedding and extraction, respectively.

Extensive experiments demonstrate that our proposed methodology does not degrade the quality of the original timbre while guaranteeing the generalization and robustness against different existing and adaptive voice cloning attacks. Besides, we verify its effectiveness with some real-world commercial services, including PaddleSpeech \cite{PaddleSpeech}, Voice-Cloning-App \cite{vca}, and so-vits-svc \cite{svs}. We also conduct some ablation studies to evaluate our design, and provide some discussions and potential insights. We believe our proposed ``Timbre Watermarking'' can shed light on the field of illegal voice cloning detection and timbre protection.

In summary, the primary contributions of our work are concluded as follows:

\begin{itemize}[leftmargin=*]
    \item We point out that timbre rights are at significant risk of being compromised by voice cloning attacks and introduce a novel concept of ``Timbre Watermarking'' as a viable defense.

    \item To achieve ``Timbre Watermarking'', we propose an end-to-end voice cloning-resistant audio watermarking framework. Innovatively, we repeatedly embed watermark information in the frequency domain to resist common audio processing and modulate the shared process of different voice cloning attacks as a distortion layer to obtain generalization.

    \item Extensive experiments demonstrate the generalization and robustness of our proposed method against different voice cloning attacks including the adaptive ones. In addition, our method is applicable in real-world services, \eg, PaddleSpeech, Voice-Cloning-App, and so-vits-svc.

\end{itemize}  

%% file: 2_background.tex
\section{Background and Related Works}
In this section, we first describe the common methods for voice cloning, followed by some countermeasures against voice cloning attacks. Finally, we provide some traditional audio watermarking methods and point out their limitations. 
\subsection{Voice Cloning} \label{sec:re-vc}
    \begin{figure}[t] 		
        \centering	
        \includegraphics[width=\linewidth]{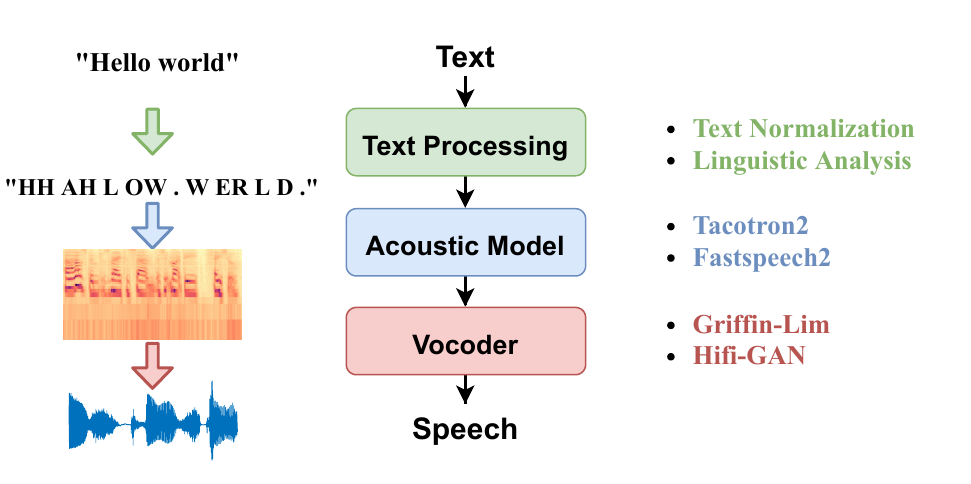}
            \vspace{-15pt}
        \caption{ The pipeline of the common TTS system.}
        \vspace{-20pt}
        \label{fig:tts}
    \end{figure}
    Voice cloning refers to the process of creating a synthetic voice that closely resembles the voice/timbre of a target person. This is mainly achieved by two mainstream techniques, namely voice conversion \cite{kaneko2019cyclegan, qian2019autovc} and text-to-speech (TTS) generation \cite{wang2017tacotron, ren2020fastspeech}. 
    Specifically, voice conversion is a technique that modifies the speech signal of an arbitrary speaker to make it sound like the target speaker's voice while preserving the linguistic content of the original message. 
    Comparably, TTS is a more flexible method to generate the desired speech of any given text without the need of the original speech for transfer. Therefore, this paper mainly focuses on the TTS-based voice cloning attacks. Nevertheless, our proposed method is also applicable to voice conversion techniques: in \Sref{sec:vc-tools} we showcase the effectiveness of our solution over real-world services, some of which are based on voice conversion. 
    \Fref{fig:tts} presents the pipeline of a common TTS system, which can be divided into the following three components.     
    
    \para{Text Processing.}
    In order to generate high-quality speech output, the text needs to be preprocessed to extract linguistic and acoustic features. Text processing for TTS mainly includes text normalization \cite{mansfield2019neural} and linguistic analysis \cite{arik2017deep}.
    Text normalization aims to convert the text into a standard form that can be processed by the TTS system. This process consists of different operations, such as removing punctuation and special characters, converting numbers to their spoken form, and expanding abbreviations and acronyms.
    Linguistic analysis segments the text into units such as sentences, phrases, and words. It then processes the normalized text by converting it into a sequence of phonemes, which are the basic units of sound in a language, and tokenizes these phonemes using the model-specific tokenizer. The resulting tokenized phonemes are then fed into the models to generate synthesized speech. Additionally, advanced speech synthesis systems often perform prosody prediction, encompassing rhythm, stress, and intonation of speech. This prediction enables TTS models to produce speech with natural pitch, duration, and energy patterns. Some TTS models integrate syntactic and semantic analysis to comprehend the text's structure and meaning, facilitating more precise and contextually relevant speech synthesis \cite{kim2021conditional}.
    
    \para{Acoustic Model.}
    An acoustic model is a key component of TTS systems, responsible for converting linguistic information into acoustic features (\eg, spectrograms), which determine the sound of the synthesized speech. Compared with traditional simple statistical models \cite{hunt1996unit, black1997automatically, tokuda2013speech, toda2007speech}, DNN-based acoustic models \cite{ping2018deep,wang2017tacotron,shen2018natural,ren2020fastspeech} give a superior performance.
        Wang \etal \cite{wang2017tacotron} proposed the first DNN-based acoustic model
        Tacotron, which uses a Recurrent Neural Network (RNN) to simulate the dynamic nature of speech signals and employs an attention mechanism to adjust the output based on different inputs. However, Tacotron has some distortion and noise issues. 
    Afterward, Shen \etal \cite{shen2018natural} presented the enhanced Tacotron 2, which adopts a position-sensitive attention module to improve the synthesis quality. Nevertheless, both Tacotron and Tacotron 2 are computationally-intensive. 
    To address the efficiency issue, Ping \etal \cite{ping2018deep} introduced a fully-convolutional sequence-to-sequence architecture rather than RNN. 
    FastSpeech \cite{ren2019fastspeech} uses an encoder-decoder transformer structure to rapidly generate mel-spectrogram in parallel for TTS and leverages an unsupervised approach to greatly simplify the model training process and reduce the demand for a large amount of audio data. 
    FastSpeech 2 \cite{ren2020fastspeech} further expands on this unsupervised training method by incorporating an acoustic prior to improve the output quality. In this paper, we adopt Tacotron 2 \cite{shen2018natural} and FastSpeech 2 \cite{ren2020fastspeech} as the default acoustic model for voice cloning attacks, but our solution is general to other models as well.
    
    \para{Vocoder.}
    With the above-obtained spectrograms, a vocoder is used to synthesize speech signals based on an analysis of their constituent frequency bands and pitch information. A well-known vocoder is Griffin-Lim \cite{griffin1984signal}, which reconstructs the original signal from Short-Time Fourier Transform (STFT). Although this algorithm is computationally inexpensive and effective at reconstructing speech signals, it will introduce perceptible background noise and distortions in the synthetic speech. It is also sensitive to the choice of parameters, which requires manual fine-tuning to achieve optimal results. Besides this traditional algorithm, there are also some deep learning algorithms used for generating high-quality audio signals, such as WaveGAN \cite{yamamoto2020parallel} and HiFiGAN \cite{kong2020hifi}. By using a combination of convolutional and deconvolutional layers, WaveGAN is able to generate realistic audio signals that mimic real recordings, but it may sometimes produce distorted or unstable output when the generated signals are too complex or too long. By comparison, HiFiGAN uses a more advanced training process that includes the use of feature-matching loss functions and a multi-scale discriminative model. This allows to generate high-fidelity audio signals that sound more natural and accurate than those generated using earlier GAN-based methods. In this paper, we adopt Griffin-Lim \cite{griffin1984signal} and HiFiGAN \cite{kong2020hifi} as the default vocoder.
    Besides, we also consider the widely-used VITS \cite{kim2021conditional}, which directly transfers the text input to speech output in an end-to-end manner.

\subsection{Countermeasures against Voice Cloning Attacks}
Existing strategies of resisting voice cloning attacks can be roughly classified into two categories.

    \para{Passive Detection-based Strategy.}
        Passive detection seeks to identify whether the suspect speech from authentic humans or artificially generated. This is normally achieved via the analysis of specific features.
        For example,
        Gao \etal \cite{gao2010audio} introduced an audio-based CAPTCHA system to differentiate human voices from synthetic ones using metrics such as short-term energy, average amplitude, and zero-crossing rate. 
        DeepSonar \cite{wang2020deepsonar}
        leverages layer-wise neuron activation patterns to effectively discern authentic voices from AI-synthesized ones.
        However, both the above methods cannot generalize to unseen data distributions. To remedy this issue, 
        Ahmed \etal \cite{ahmed2020void} proposed Void, an efficient solution for voice spoofing detection based on liveness detection, which analyzes the spectral power differences between live-human voices and spoofing voices.
        In addition to feature-based methods, there are also some approaches to achieve good detection capability with the help of deep learning \cite{lieto2019hello,li2021replay}.  
        To solve the problem of diverse statistical distributions of different synthesizing methods, Zhang \etal \cite{zhang2021one} proposed a one-class learning anti-spoofing system (One-Class) to effectively detect unseen synthetic voice spoofing attacks.

        While these passive detection methods can  identify synthetic speech in specific scenarios (\eg particular datasets and contexts), their generalizability and credibility are limited  \cite{wenger2021hello}. Most importantly, they can only detect synthetic speech, but not trace the original timbre. We showcase the corresponding comparison in \ASref{sec:det}.

    \para{Proactive Prevention-based Strategy.}
    This line of solutions focus on disrupting the voice cloning process rather than detecting the synthesized speech post facto.
        Huang \etal \cite{huang2021defending} proposed to corrupt speech samples by adding adversarial perturbations to prevent unauthorized speech synthesis. However, it will degrade the quality of the target speech. Besides, it can only be applicable for voice conversion tasks at the inference stage, rather than the complex training process of TTS tasks. Nevertheless, we compare our watermarked speech with their perturbated one to demonstrate our superior performance on speech quality in \Sref{sec:experiment-fidelity}.

    \begin{figure}[t] 		
        \centering	
        \includegraphics[width=\linewidth]{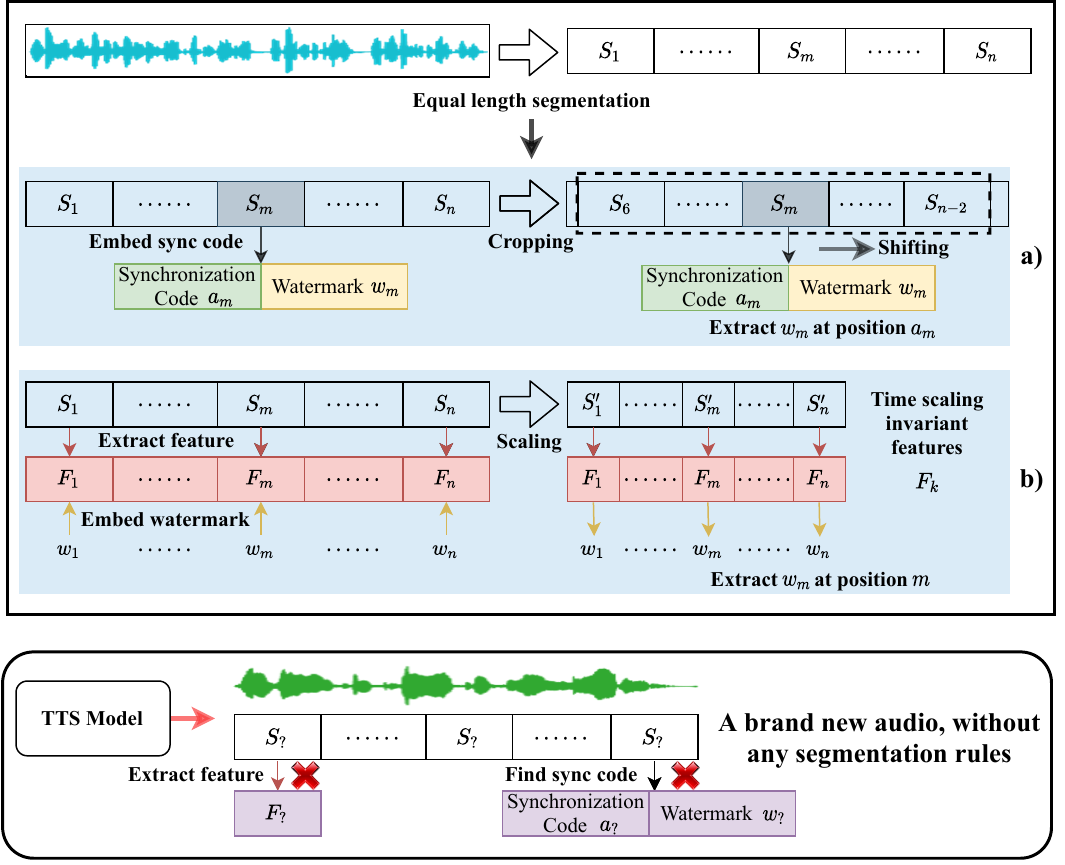}
        \caption{Traditional audio watermarking schemes are unable to withstand voice cloning attacks.}
        \vspace{-20pt}
        \label{fig:traditional-audio-wm}
    \end{figure}
    
\subsection{Audio Watermarking} \label{sec:audio-wm-background}

   \hl{Audio watermarking technologies strive to achieve a dual set of performance criteria: fidelity and robustness. For fidelity, current audio watermarking methods predominantly utilize frequency domain analysis to extract salient features from the carrier audio. The embedding of the watermark is subsequently executed by altering these frequency domain features. It is generally accepted that modification of the higher frequency components leads to enhanced fidelity in the watermarking process \cite{lin2015audio}. Concurrently, making subtle adjustments to less prominent audio features contributes to the watermark's imperceptibility \cite{su2018snr, zhao2022ssvs}. Besides, some schemes try to use psychoacoustic modeling to set thresholds for watermark modifications to avoid being perceived by human ears \cite{lie2006robust,lee2000digital, swanson1998robust}. For robustness, some research has demonstrated resilience against common and unavoidable signal processing distortions \cite{hua2016twenty}. More recently, the focus has shifted toward exploring robustness under intricate conditions, such as re-recording \cite{liu2018patchwork, liu2022dear}, desynchronization distortion \cite{zhao2021desynchronization, zhao2022ssvs}, cropping distortion \cite{zhang2023m}, etc.
   In a nutshell, existing audio watermarking can already well balance the trade-off between fidelity and normal robustness.
   }

    As mentioned above, the collected speech may be modified before voice cloning attacks via some preprocessing operations, such as cropping and time scaling. This can cause desynchronization to fail the subsequent watermark extraction.     
    \Fref{fig:traditional-audio-wm} illustrates some countermeasures based on traditional audio watermarking schemes: a) leveraging synchronization codes or shift mechanisms with sliding windows for resynchronization \cite{liu2018patchwork, zhang2023m}; b) employing time scaling invariant features to withstand scale transformations \cite{zhao2021desynchronization,zhao2022ssvs}. 
    Based on these two ideas, Zhao \etal \cite{zhao2021desynchronization} designed FSVC, an audio feature that is insensitive to time-domain scale transformation based on frequency-domain signal processing. It is combined with audio averaging segmentation to achieve resistance to desynchronization caused by global-scale transformation.  Liu \etal \cite{liu2018patchwork} designed RFDLM, a robust feature against scaling. It is combined with synchronization codes to achieve robustness to desynchronization attacks.
    \hl{However, \textit{none of these audio watermarking methods
    exhibit satisfactory robustness against both cropping and time scaling}, let alone the more complex voice cloning process.} 
    In \Sref{sec:appendix-comparison} we evaluate two audio watermarking methods (\ie, FSVC \cite{zhao2021desynchronization} and RFDLM \cite{liu2018patchwork}), and show that they \textit{totally fail} to detect voice cloning attacks. 
    {
    \hl{
    Concurrently, deep learning-based end-to-end audio watermarking has recently emerged \cite{liu2022dear,pavlovic2022robust,wang2021generating}, yet it encounters considerable limitations. Existing methods prioritize robustness; however, in the given scenario, the audio possesses a fixed length, rendering it challenging to counteract distortions arising from desynchronization.}
    }

    Different from the above methods, our solution integrates time-independent features into the audio watermarking algorithm, making it feasible to maintain the integrity of the watermark information even in the face of cropping and time scaling. We provide a qualitative analysis of the impact of distortion operations on the watermarked speech and demonstrate that our method can effectively resist voice cloning attacks. 
\hl{
In a nutshell, our method pursues a novel robustness property against voice cloning attacks, which is achieved by simulating this attack and inserting it between the training of watermark embedding and extraction. For enhanced fidelity, we also introduce an extra discriminator for adversarial training. 
}
    

%% file: 3_preliminary.tex

\section{Threat Model} \label{sec:tm}

We consider a scenario which involves three parties: 1) \textbf{users} shares their audio data on a public platform; 2) \textbf{the platform provider} seeks to protect the shared audio from potential misuse; 3) \textbf{the attacker} collects the target audio from the platform and attempts to generate synthetic audio with the same timbre using advanced voice cloning techniques.

\subsection{Users' Ability and Goal}
In order to access the platform, users must first register the service with their authorship information. They hope that their audio data will not be used maliciously. They have the right to request the platform to deploy protection over the uploaded audio, and perform synthetic audio verification when suspiciously cloned audio is identified.

\subsection{Platform Provider's Ability and Goal}

The platform provider applies a novel watermark embedding algorithm $\mathcal{EM}(\cdot)$ to embed a watermark $w$ (\eg, authorship information)
    into users' speech samples $s$ before releasing them, to safeguard their voice timbre, \ie, $s^{\prime} = \mathcal{EM}(s,w)$. The platform provider needs to guarantee the speech \textbf{fidelity}, namely, making $s_w$ similar to the pristine $s$ as much as possible. Let $S_w = \{s_{1_w}, s_{2_w}, \dots, s_{n_w}\}$ represent a set of watermarked speech samples from the target speaker $T$ with watermark $w$. The platform provider also has a watermark extraction algorithm $\mathcal{EX}(\cdot)$ to extract the pre-defined watermark $w$ from $s_w$, \ie,  $w' = \mathcal{EX}(s_w) \rightarrow w$. Given a suspicious audio $s*$,  the platform provider can apply $\mathcal{EX}(\cdot)$ to check if any watermark can be extracted from it, as evidence of voice cloning attacks.

\begin{figure*}[t] 		
    \centering	
    \includegraphics[scale=0.65]{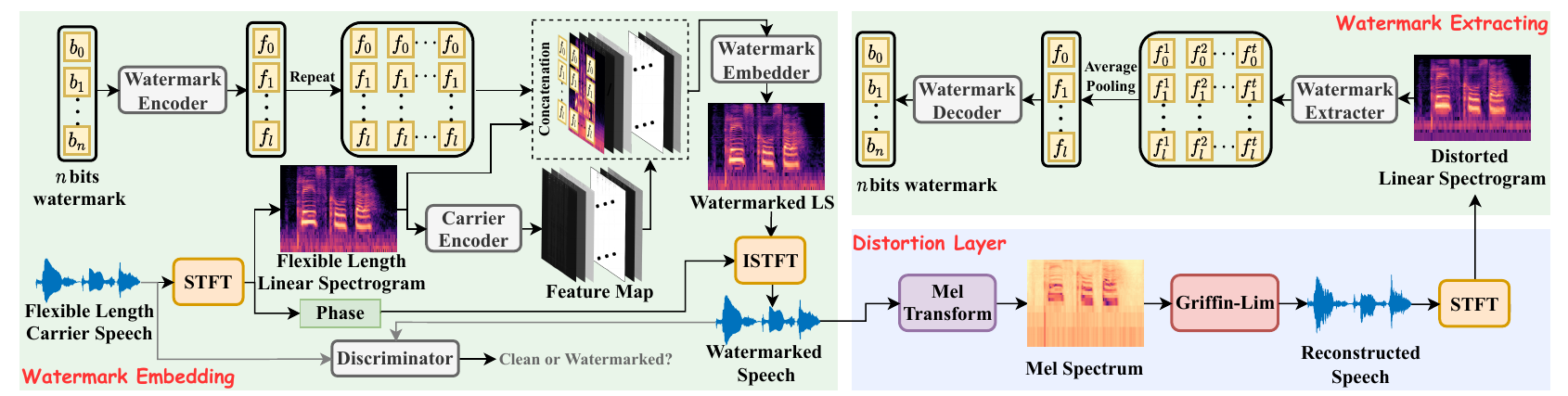}
    
    \caption{Overview of our proposed timbre watermark framework.}
    \vspace{-20pt}
    \label{fig:framework}
\end{figure*}

\subsection{Attacker's Ability and Goal} \label{sec:att-ability}
\label{sec:threatmodel-attacker}

    Let $S = \{s_1, s_2, ..., s_n\}$ be a set of speech samples from a target speaker $T$. The attacker collects $S$ and pre-processes it to construct a text-speech dataset
    $D=\{(x_1,y_1), (x_2,y_2), ..., (x_m,y_m)\}$, where $x_i$ is a text transcription and $y_i$ is the corresponding speech sample. With $D$,  the attacker can train his TTS model $\mathbf{M}$ in a supervised way. As mentioned in \Sref{sec:re-vc}, a TTS model $\mathbf{M}$ usually consists of two components, namely, the Acoustic model $\mathbf{M_A}$ and the Vocoder $\mathbf{M_V}$. After that, for an arbitrary text transcription $\hat{x}$, $\mathbf{M}$ can synthesize the corresponding speech $\hat{y}$ as:
    $$\hat{y} = \mathbf{M}(\hat{x}) = \mathbf{M_V}(\mathbf{M_A}(\hat{x})).$$

    With our proposed Timbre Watermark, the attacker can only obtain a watermarked dataset from the platform, \ie,  
    $D_w = \{(x_1, y_{1_w}), (x_2, y_{2_w}), \dots, (x_{m}, y_{m_w})\}$, where $y_{i_w}$ denotes a watermarked speech sample. 
    Then he uses $D_w$ to train his acoustic model $\mathbf{M_{A_w}}$. Based on the attacker's capability of training the subsequent vocoder $\mathbf{M_V}$, we define the following three attack scenarios:

\begin{itemize}[leftmargin=*]

    \item \textit{Professional Voice Cloning Attack.} In this 
    scenario, the attacker is an expert on TTS tasks, and has the capability of fine-tuning a general vocoder $\mathbf{M_V}$ on $D_w$ to acquire his superior vocoder $\mathbf{M_{V_w}}$.
    This vocoder $\mathbf{M_{V_w}}$ can convert the synthesized mel-spectrograms into counterfeit speech closely resembling the target speaker's timbre. The attack scenario can be formulated as follows:
    $${\hat{y_w}}^P = {\mathbf{M_w}}^P(\hat{x}) = \mathbf{M_{V_w}}(\mathbf{M_{A_w}}(\hat{x})).$$
    
    \item \textit{Regular Voice Cloning Attack.} 
    In this scenario, the attacker only trains his own acoustic model $\mathbf{M_{A_w}}$ and then appends an off-the-shelf pre-trained vocoder $\mathbf{M_{V}}$ to conduct voice cloning attacks, \ie, 
    $${\hat{y_w}}^R = {\mathbf{M_w}}^R(\hat{x}) = \mathbf{M_{V}}(\mathbf{M_{A_{w}}}(\hat{x})).$$
    
    \item \textit{Low-quality Voice Cloning Attack.} In certain instances, the attacker may lack the resources necessary to obtain and utilize high-quality pre-trained vocoders for speech synthesis, or the capability of fine-tuning vocoders to target their victims specifically. Consequently, he might resort to traditional, non-deep learning techniques for synthesizing speech waveforms, such as the Griffin-Lim algorithm $\mathbf{GL(\cdot)}$ \cite{griffin1984signal}. 
    The attack scenario can be described as follows:
    $${\hat{y_w}}^L = \mathbf{M_w}^L(x) = \mathbf{GL}(\mathbf{M_{A_{w}}}(x)).$$

\end{itemize}

\subsection{Adaptive Attacks}
\label{sec:adp}
Besides the above common voice cloning attacks, we also consider the possible adaptive attacks, where the attacker has the knowledge of our timbre watermarking strategy, and tries to remove the watermarks while preserving the quality of the cloned audio. We implement and evaluate the following adaptive attacks in \Sref{sec:experiment-adaptive}:

\hl{
\noindent\textbf{1) Attackers without access to the watermarking model.} The attacker is not allowed to access the watermarking model and can only launch some data-level attacks as follows:
\begin{itemize}[leftmargin=*]
\item Before conducting the voice cloning attack, the attacker preprocesses the collected audio with some harmful operations, such as severe compression and low pass filter.
\item The attacker can launch the watermark evading attack with VAE reconstruction like \cite{zhao2023generative} before the voice cloning attack.
\item The attacker can collect some pristine data of the target speaker, and synthesize the audio with the mixed dataset.
\end{itemize}}

\hl{
\noindent\textbf{2) Attackers with access to the watermarking model.} When an attacker can access the target model, he has more means to break the watermark as follows:
\begin{itemize}[leftmargin=*]
\item With access to the watermark encoder, the attacker can operate a watermark overwriting attack, \ie, further embedding another watermark on the collected watermarked data before the voice cloning attack.
\item The attacker can execute the watermark evading attack by deploying a watermark erasing VAE before voice cloning.
\item With access to the watermark extractor, the attacker attempts to explore the location of the embedded watermark, and then directly remove this region before applying voice cloning.
\item The attacker can also deploy a classifier to detect the presence of a watermark. This classifier is then used against a speech synthesis model during training (\ie, domain-adversarial training), which causes the synthesis model to synthesize audio without the watermark.
\end{itemize}}
\hl{
\noindent\textbf{3) Combining multiple attack strategies.}
We consider integrating diverse
attack schemes, encompassing regular preprocessing, harmful preprocessing, domain-specific advanced training, VAE reconstruction, and
watermark overwriting.}

%% file: 4_method.tex
\section{Methodology} \label{sec:method}

\Fref{fig:framework} shows the overview of our new framework. It consists of three components: a watermark embedding module, a watermark extraction module, and an intervening distortion layer to bolster the robustness against distortions. These components are jointly trained. Below we provide a detailed description of each component.

\subsection{Watermark Embedding} \label{sec:method-embed}
\hl{
    As shown in the left part of \Fref{fig:framework}, similar to many audio-based information hiding methods \cite{ozer2005svd, kreuk2019hide, jiang2020smartsteganogaphy}, we adopt the widely-used linear spectrogram \cite{itakura1975line} of speech audios as the carrier to embed watermark information. 
    Meanwhile, this step allows us to add the same watermark in the frequency domain at different time periods for time-independence, due to the short time window property brought about by the time-frequency localization of STFT.
}
Specifically, given a single-channel raw speech audio $a$ of the flexible length $N$, we first apply the Short-Time Fourier Transform operation ($\mathbf{STFT}(\cdot)$) on it to produce a spectrogram $s$ and the corresponding phase information $p$:
\begin{equation}
s, p = \mathbf{STFT}(a).
\label{eq:stft}
\end{equation}
\hl{
Since the magnitude spectrogram contains most of the information, we use the magnitude spectrogram $s$ as the carrier for easy training, while the phase spectrogram is only used for signal recovery \cite{ozer2005svd,kreuk2019hide}.
}
Then, we feed $s$ into the Carrier Encoder $\mathbf{EN_{c}}$ to obtain the encoded carrier features $f_{c}$:
\begin{equation}
f_{c} = \mathbf{EN_{c}}(s).
\end{equation}
Simultaneously, we feed the $n$-bit watermark information $w$ into the Watermark Encoder $\mathbf{EN_{w}}$ to obtain the encoded watermark features $f_{w}$:
\begin{equation}
f_{w} = \mathbf{EN_{w}}(w).
\end{equation}
%
Next, we concatenate $f_{c}$ and $f_{w}$ to obtain the final input $f_{+}$ for the subsequent Watermark Embedder $\mathbf{EM}$.
Considering that the length of speech samples is flexible, we propose to repeat $f_{w}$ along the time axis
\hl{
 to achieve time-independence
}
, \ie, $\text{Repeat}(f_{w}, t)$, whose shape size is consistent to $f_{c}$. We point out that such a repeated strategy can make the watermark information inherently robust to distortions in the time domain.  
\hl{
Motivated by DenseNet \cite{huang2017densely}, we also introduce a skip concatenation to increase the nonlinearity
and characterization ability of the model and preserve the information of the original carrier $s$ to the maximum extent.
All this preprocessing can be written as follows:
}
\begin{equation}
f_{+} = \mathbf{Concatenate}(f_{c}, s, \text{Repeat}(f_{w}, T)),
\label{eq:repeat}
\end{equation}
where $f_{w}\in \mathbb{R}^{C_w\times 1\times H}$, $f_{c}\in \mathbb{R}^{C_v\times T\times H}$, and $f_{+}\in \mathbb{R}^{(C_w + 1 + C_v)\times T\times H}$ .
Afterward, $f_{+}$ is fed into $\mathbf{EM}$ to obtain the watermark embedded spectrogram $s_w$:
\begin{equation}
s_w = \mathbf{EM}(f_{+}).
\end{equation}
Finally, we reconstruct the watermarked audio $a_w$ by applying Inverse Short-Time Fourier Transform ($\mathbf{ISTFT}(\cdot)$) to the decoded spectrogram $s_w$ and original phase information $p$: 
\begin{equation}
a_w = \mathbf{ISTFT}(s_w, p).
\label{eq:istft}
\end{equation}

To ensure the audio quality fidelity, we introduce the watermark embedding loss $\mathcal{L}_{e}$, \ie, 
\begin{equation}
\mathcal{L}_{e} = \frac{1}{M}\sum_{i=1}^{M}((a_w)_{i}-a_{i})^2,
\end{equation}
where $M$ is the length of the audio in the time dimension.
To further improve the fidelity and minimize the domain gap between $a$ and $a_w$, 
\hl{
inspired by the training strategy in GAN \cite{goodfellow2020generative} to ensure the realism of the generated data, 
an adversarial loss $\mathcal{L}_{adv}$ is also added to make an extra discriminator $\mathbf{D}$ cannot distinguish $a_w$ from the pristine $a$, as shown in \Fref{fig:framework},} \ie,
\begin{equation}
    \mathcal{L}_{adv} = -\log(\sigma(\mathbf{D}(a_w))).
\end{equation}
Meanwhile, $\mathbf{D}$ is constrained by $\mathcal{L}_{d} = -\log(\sigma(\mathbf{D}(a))) -\log(1 - \sigma(\mathbf{D}(a_w)))$, where $\sigma(\cdot)$ denotes the sigmoid function.

\subsection{Watermark Extracting}
Given a watermarked speech $a_w$, the decoder needs to recover watermark $w^{\prime}$ as consistent as the original watermark $w$. 
Specifically, we first follow \Eref{eq:stft} to conduct STFT on $a_w$ to obtain the phase information $p_w$ and spectrogram $s_w$, which are then fed into the Watermark Extractor $\mathbf{EX}$ to obtain the extracted watermark features $f_{w}^{\prime}$:
\begin{equation}
f_{w}^{\prime} = \mathbf{EX}(s_w).
\end{equation}
Then, we use the Watermark Decoder $\mathbf{DE}$ to decode the watermark information $w^{\prime}$ from the time domain (horizontal direction) averaged watermark features $f_{w}^{\prime}$, \ie,
\begin{equation}
w^{\prime} = \mathbf{DE}(\mathbf{Average}(f_{w}^{\prime})).
\end{equation}
\hl{
This average operation corresponds to the previous repeat operation in \Eref{eq:repeat}, and together they realize the time-independence of watermarking.
}
To ensure the accuracy of watermark extraction, we introduce a watermark extraction loss $\mathcal{L}_{w}$, \ie, 
\begin{equation}
\mathcal{L}_{w} = \frac{1}{N}\sum_{i=1}^{N} (w_{i}^{\prime}- w_{i})^2,
\end{equation}
where $N$ is the length of the watermark sequence.

\subsection{Distortion Layer} \label{sec:method-dl}
To enhance the robustness against voice cloning attacks,
we further insert a distortion layer between the watermark embedding stage and watermark extracting stage. Given a watermarked audio signal $a_w$, we can obtain the distorted audio $\hat{a_w}$ after the distortion layer $\mathbf{DP}$, \ie, $\hat{a_w} = \mathbf{DP}(a_w)$, which is then fed into the watermark extracting stage. Note that the distortion layer is only appended in the training stage, and will be discarded in the actual embedding and extraction stages.
Below we detail each component of the distortion layer.

{

\para{\hl{ISTFT Distortion.}}
\hl{
When performing ISTFT, if the amplitude spectrogram is modified in some way to embed the watermark while the phase spectrogram remains unchanged, then the final signal obtained will be a complex matrix, and we will only keep the real part as the final signal, which incurs some loss. This differential distortion is already included in the embedding process as shown in \Eref{eq:istft}.
}

\para{Normalization Distortion.}
During speech synthesis training, the timbre information is unrelated to the overall energy, or amplitude, of the audio signal. Consequently, the speech synthesis model often normalizes the amplitude of the audio as part of the preprocessing, \ie,
\begin{equation}
a_w^n = \frac{a_w}{\max(|a_w|)}.
\end{equation}
This normalization operation, however, may result in the potential erasure of the watermark information embedded within the audio signal. Thus, we append it into the distortion layer.

\para{Transformation Distortion.}
For speech synthesis, mel-spectrogram is mainly used as supervisory signals. Specifically, during model training, the audio needs to be mel-transformed to extract this feature from the speech signal. Therefore, in the distortion layer, we perform the mel-transform operation on the watermarked audio to derive a mel-spectrogram,
\ie,
\begin{equation}
\hat{{ms}}_w = \mathbf{Mel}(a_w^n).\\
\end{equation}

\para{Wave Reconstruction Distortion.}
During the synthesis process, the vocoder is employed to reconstruct a waveform from a mel-spectrogram while omitting the phase information. This transformation is irreversible and lossy, which can affect the intermediate audio features, particularly the spectrogram, and potentially lead to the loss of embedded watermark information. Consequently, the resulting distribution deviates from its original one. To address this issue, we employ the vocoder for waveform reconstruction. 

Specifically, we adopt the conventional rule-based Griffin-Lim \cite{griffin1984signal} $\mathbf{GL}(\cdot)$ as the vocoder, 
\ie,
\begin{equation}
\hat{a_w} = \mathbf{GL}(\hat{{ms}}_w). 
\end{equation}
\hl{This operation introduces more severe distortions compared to learnable vocoders, and we assume it can enhance the transferable robustness (\Tref{tb:vc} supports our assumption). } 

\para{Pipeline of the Distortion Layer.}
The pipeline of the distortion layer $\mathbf{DP}$ can be written as follows:
\hl{
\begin{equation}
    \mathbf{DP}(a_w) = \mathbf{GL}(\mathbf{Mel}(\frac{a_w}{\max(|a_w|)})).
\end{equation}
}

Given the distorted watermarked speech $\hat{a_w}$, the decoder needs to recover the watermark $\hat{w^{\prime}}$ as consistent as the original watermark $w$. To recover the watermark information $\hat{w^{\prime}}$ from $\hat{a_w}$, we first apply STFT on $\hat{a_w}$ to obtain the phase information $\hat{p_w}$ and spectrogram $\hat{s_w}$, which are then fed into the Watermark Extractor $\mathbf{EX}$ to obtain the extracted watermark features $\hat{f_{w}^{\prime}}$:
\begin{equation}
\hat{f_{w}^{\prime}} = \mathbf{EX}(\hat{s_w}).
\end{equation}
Similarly, we can get the recovered watermark $\hat{w^{\prime}}$ from $\hat{f_{w}^{\prime}}$ using the Watermark Decoder $\mathbf{DE}$, \ie, 
\begin{equation}
\hat{w^{\prime}} = \mathbf{DE}(\mathbf{Average}(\hat{f_{w}^{\prime}})).
\end{equation}
Here, we introduce $\hat{\mathcal{L}_{w}}$ to ensure the accuracy of watermark extraction after the distortion layer, 
\ie,
\begin{equation}
\hat{\mathcal{L}_{w}} = \frac{1}{m}\sum_{i=1}^{m} (\hat{w^{\prime}}_i- w_{i})^2.
\end{equation}

\subsection{End-to-end Protection} \label{sec:method-train}

\para{Model Training.}
We jointly train the above three modules in our framework.
The whole loss function $\mathcal{L}$ is formulated as: 
\begin{equation}
\label{eq:final-loss}
    \mathcal{L} = \lambda_e \cdot \mathcal{L}_{e} + \lambda_{adv} \cdot \mathcal{L}_{adv} + \lambda_w \cdot (\mathcal{L}_{w}+\hat{\mathcal{L}_{w}}),
\end{equation}
where $\lambda_e$, $\lambda_d$, and $\lambda_w$ are hyper-parameters to balance the three terms. \hl{It is worth mentioning that we simultaneously optimize for watermark extraction accuracy from both distorted watermarked audio and undistorted watermarked audio, ensuring generalization.} At the same time, we train $\mathbf{D}$ to minimize the loss $\mathcal{L}_{d}$.

\para{Watermark Embedding.}
As shown in \Fref{fig:f1}, before releasing the original audio, the platform provider embeds the pre-defined watermark information (\eg, a bit string) into it by the well-trained $\mathbf{EN_{c}}$, $\mathbf{EN_{w}}$, and $\mathbf{EM}$.

\para{Watermark Extraction and Attack Detection.}
Given a suspicious audio, we adopt the well-trained Extractor $\mathbf{EX}$ and $\mathbf{DE}$ to extract the watermark from it. Then, we use bit recovery accuracy to evaluate the similarity between the extracted and ground-truth watermark. If the accuracy is larger than the pre-defined threshold, we can finally claim that this suspicious audio is synthesized from the protected ones without permission, and the voice cloning attack is detected.

%% file: 5_experiments.tex
\section{Experiments}
\label{sec:experiment}
We present comprehensive evaluations to validate the effectiveness of our framework. We first describe our experiment setup in \Sref{sec:setting}. Then we demonstrate the proposed framework can satisfy different requirements, including fidelity (\Sref{sec:experiment-fidelity}), generalization (\Sref{sec:robustvc}), robustness against potential distortions (\Sref{sec:common}) as well as adaptive attacks (\Sref{sec:experiment-adaptive}). In \Sref{sec:vc-tools} we conduct evaluations on three real-world services to show the practicality of our solution. \Sref{sec:ablation-study} gives the ablation study to verify our design. 

\subsection{Experiment Setup} \label{sec:setting}

\para{Implementation Details.} \label{sec:detail}
\hl{
For all of $\mathbf{EN_{c}}$, $\mathbf{EM}$, and $\mathbf{EX}$, we adopt simple fully 2D convolutional networks. 
These networks maintain the dimensions of feature maps at each layer and employ a skip gated block as their fundamental unit, which integrates the Gated Convolutional Neural Network \cite{dauphin2017language} with a skip connection \cite{he2016deep} for processing. For Watermark Encoder $\mathbf{EN_{w}}$, we leverage a fully connected layer that utilizes the LeakyReLU activation function \cite{xu2015empirical} for enhanced performance. Meanwhile, the Watermark Decoder, denoted as $\mathbf{DE}$, operates as a pure linear layer within the network architecture. The Discriminator $\mathbf{D}$ has a comprehensive architecture consisting of STFT, three ReluBlocks, an average pooling layer, and a linear layer. The ultimate output of this structure is the prediction result for the input audio. The implementation of the models is available through the source code in our website.
}
Additionally, to jointly train the three modules in our platform, we use the Adam optimizer \cite{kingma2014adam} with $\beta_1=0.9$, $\beta_2=0.98$, $\epsilon=10^{-9}$, and a learning rate of $2e^{-5}$ for optimization. We set $\lambda_e=1$ and $\lambda_w=\lambda_d=0.01$ in Eq.(\ref{eq:final-loss}). For STFT, we adopt a filter length of $1024$, a hop length of $256$, and a window function applied to each frame with a length of $1024$.

\para{Datasets.} \label{sec:dataset}
For the voice cloning model, we employ LJSpeech \cite{ljspeech17} (Clip length varies from 1 to 10 seconds), a well-established benchmark dataset for speech synthesis. It includes many well-aligned text-speech pairs.
For watermarking model training, we employ the standard training set \hl{\texttt{train\_clean100}} of LibriSpeech \cite{panayotov2015librispeech}, where the length of audio samples varies, typically around 10 seconds in duration. We also conduct testing based on its standard testing set with 2620 audio samples.
All audio samples are resampled to the sampling rate of 22.05 kHz, for both LJSpeech and LibriSpeech datasets. \hl{There is \textit{no overlap} between the training data of the watermarking model and voice cloning models.}

\para{Metrics.}
For fidelity evaluation, \ie, the quality of watermarked audio, we adopt three objective metrics:
Signal-to-Noise Ratio (${\textbf{SNR}}$), Perceptual Evaluation of Speech Quality (${\textbf{PESQ}}$) \cite{rix2001perceptual}, and Speaker Encoder Cosine Similarity (${\textbf{SECS}}$) \cite{casanova2021sc}, and an additional subjective metric: Mean Opinion Score (${\textbf{MOS}}$).
In detail, SNR is only used to measure the magnitude of quality loss resulting from watermarking and audio processing, while PESQ provides a better assessment of speech quality (\ie, imperceptibility) than SNR by considering the specifics of the human auditory system.  
SECS leverages the speaker similarity to evaluate the fidelity, and a higher value indicates a stronger similarity. We follow prior works \cite{casanova2021sc, choi2020attentron} to compute this metric using the speaker encoder of the Resemblyzer \cite{jemine2019real} package.
In practice, SECS is often used for voice authentication, and the audio is determined to bypass authentication when SECS $>0.9$ \cite{cooper2020zero}.
In MOS evaluations, we invite 10 participants to rate the naturalness and quality of the target speech with five ratings
(1: Bad, 2: Poor, 3. Fair, 4: Good, 5: Excellent).

For the effectiveness of watermark extraction, we use the bit recovery accuracy (${\textbf{ACC}}$), calculated from all audios in the standard test set of LibriSpeech (each audio is embedded with a random watermark). 
Furthermore, we evaluate the robustness against voice cloning attacks by examining ACC derived from 500 synthesized speech samples (an identical watermark), which correspond to 500 text segments from the LJSpeech test set. The default watermark length is 10 bits.

\subsection{Fidelity}
\label{sec:experiment-fidelity}
We first compare the fidelity of the protected audio between our method and the adversarial perturbation-based method \cite{huang2021defending}. It is worth noting that this method \cite{huang2021defending} is not effective in defeating different voice cloning attacks, so the comparison here is just for fidelity evaluation.  We directly use their released speech examples on the web page\footnote{\url{https://yistlin.github.io/attack-vc-demo/}} for comparison.

\begin{figure}[t] 		
    \centering	
    \includegraphics[scale=0.3]{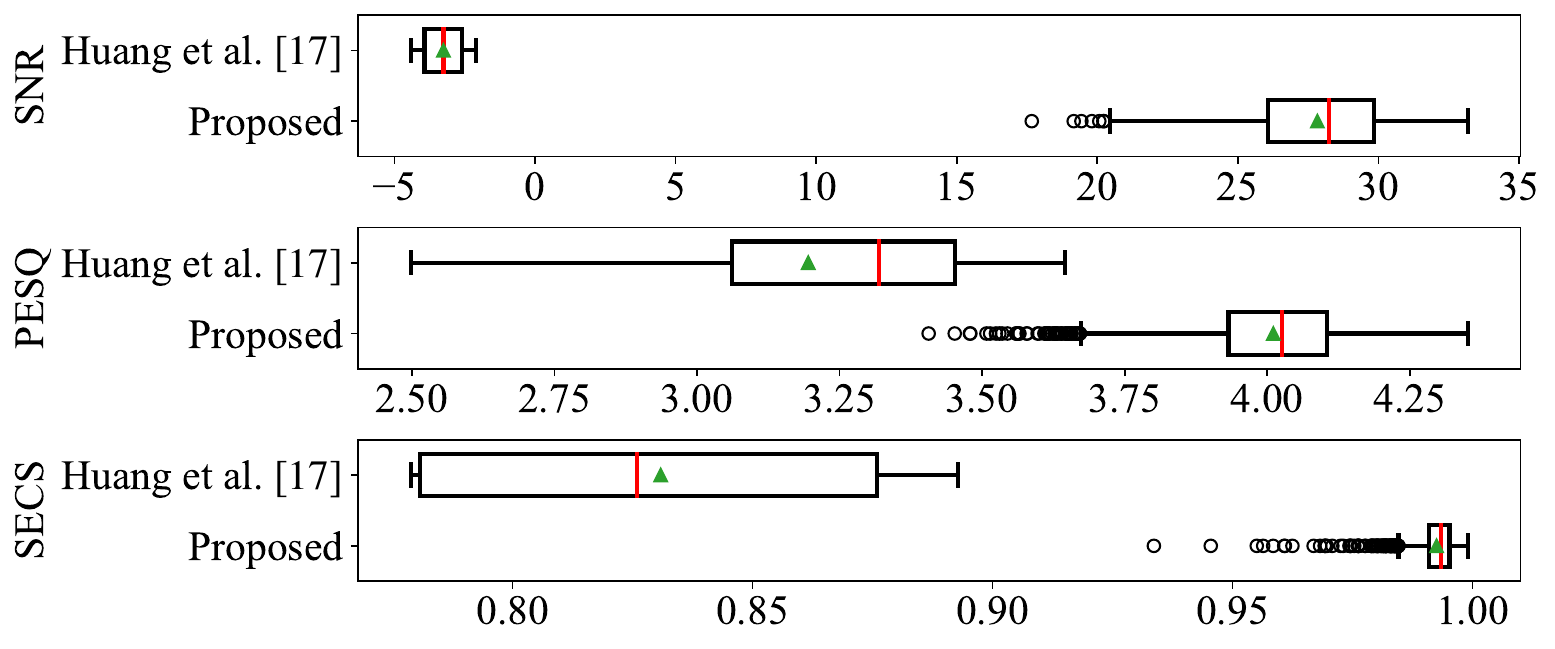}
    \vspace{-0.6em}
    \caption{Objective fidelity comparison with the baseline \cite{huang2021defending}. Green triangles represent the mean values and red lines indicate the median values.}
    \vspace{-10pt}
    \label{fig:fidelity}
\end{figure}

\begin{figure}[t] 		
    \centering	
    \includegraphics[scale=0.3]{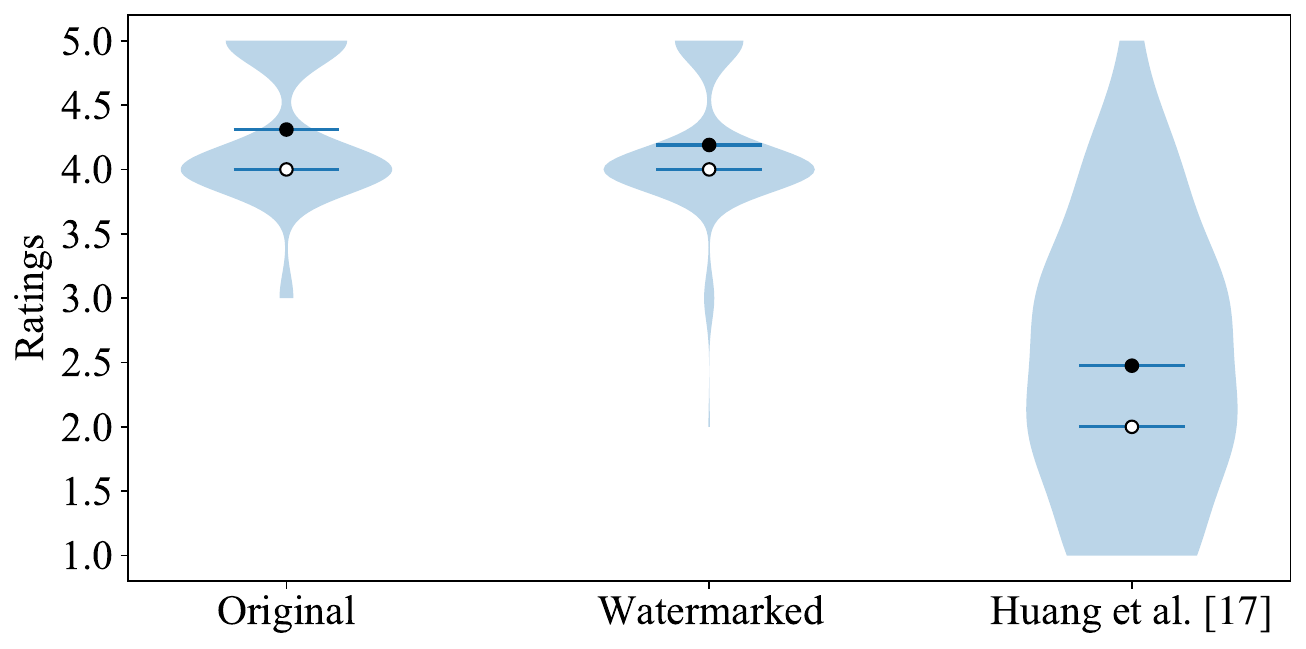}
    \vspace{-0.6em}
    \caption{Subjective fidelity comparison (MOS) with the baseline \cite{huang2021defending}. Black dots are means and white dots are medians.}
    \vspace{-15pt}
    \label{fig:mos}
\end{figure}

\Fref{fig:fidelity} shows the SNR, PESQ and SECS metrics of the two methods, respectively. For all cases, our method has a superior fidelity to Huang's method \cite{huang2021defending}. Specifically for SECS, where a lower value indicates better speaker similarity, our method outperforms \cite{huang2021defending} by a large margin. We also conduct a subjective fidelity comparison: we select 20 distinct segments of both watermarked and non-watermarked audio samples, as well as the adversarial audio specimens delineated in \cite{huang2021defending}. \Fref{fig:mos} shows the speech quality ratings from 10 participants. The conclusion is consistent with the objective evaluation. 
The corresponding audio samples for the above comparison are available on our project website. 

\begin{table}[t]
\begin{center}
\caption{Generalization across different voice cloning attacks. * denotes using a watermarked dataset to train the acoustic model or fine-tune the vocoder, otherwise using a watermark-free dataset. Red, blue, and gray areas represent professional, regular, and low-quality attacks, respectively. The quality of the related synthesized speech is also provided.}
\vspace{-0.6em}
\begin{tabular}{ccccc}
\hline
\multicolumn{2}{c}{Model}                                            & \multicolumn{2}{c}{Quality}                                     &                                \\ \cline{1-2} \cline{3-4}
Acoustic   Model                    & Vocoder                             & PESQ$\uparrow$                          & SECS$\uparrow$                        & \multirow{-2}{*}{ACC$\uparrow$}          \\ \hline
                               & \cellcolor[HTML]{FFCCC9}Hifi-GAN* \cite{kong2020hifi}   & \cellcolor[HTML]{FFCCC9}1.0578 & \cellcolor[HTML]{FFCCC9}0.8957 & \cellcolor[HTML]{FFCCC9}1.0000 \\
                               & \cellcolor[HTML]{DAE8FC}Hifi-GAN \cite{kong2020hifi}    & \cellcolor[HTML]{DAE8FC}1.0712 & \cellcolor[HTML]{DAE8FC}0.8965 & \cellcolor[HTML]{DAE8FC}0.9933 \\
\multirow{-3}{*}{Fastspeech2* \cite{ren2020fastspeech}} & \cellcolor[HTML]{EFEFEF}Griffin-Lim \cite{griffin1984signal} & \cellcolor[HTML]{EFEFEF}1.1129 & \cellcolor[HTML]{EFEFEF}0.7034 & \cellcolor[HTML]{EFEFEF}1.0000 \\ \hline
                               & \cellcolor[HTML]{FFCCC9}Hifi-GAN* \cite{kong2020hifi}   & \cellcolor[HTML]{FFCCC9}1.1143 & \cellcolor[HTML]{FFCCC9}0.8598 & \cellcolor[HTML]{FFCCC9}1.0000 \\
                               & \cellcolor[HTML]{DAE8FC}Hifi-GAN \cite{kong2020hifi}    & \cellcolor[HTML]{DAE8FC}1.1136 & \cellcolor[HTML]{DAE8FC}0.8626 & \cellcolor[HTML]{DAE8FC}0.9988 \\
\multirow{-3}{*}{Tacotron2* \cite{shen2018natural}}   & \cellcolor[HTML]{EFEFEF}Griffin-Lim \cite{griffin1984signal} & \cellcolor[HTML]{EFEFEF}1.1971 & \cellcolor[HTML]{EFEFEF}0.7125 & \cellcolor[HTML]{EFEFEF}1.0000 \\ \hline
\rowcolor[HTML]{FFCCC9} 
\multicolumn{2}{c}{\cellcolor[HTML]{FFCCC9}VITS* \cite{kim2021conditional}}                    & 1.0342                         & 0.9085                         & 1.0000                         \\ 
\hline 
\end{tabular}
\label{tb:vc}
\vspace{-10pt}
\end{center}
\end{table}

\begin{table}[t]
\begin{center}
\caption{Extraction results with all models trained on watermark-free data. }
\vspace{-0.6em}
\begin{tabular}{ccccc}
\hline
\multicolumn{2}{c}{Model}                                            & \multicolumn{2}{c}{Quality}                                     &                                \\ \cline{1-2} \cline{3-4}
Acoustic   Model                    & Vocoder                             & PESQ$\uparrow$                          & SECS$\uparrow$                        & \multirow{-2}{*}{ACC$\uparrow$}          \\ \hline
                               & Hifi-GAN \cite{kong2020hifi}                            & 1.0416                         & 0.9031                         & 0.5270                         \\
\multirow{-2}{*}{Fastspeech2 \cite{ren2020fastspeech}}  & Griffin-Lim \cite{griffin1984signal}                        & 1.0756                         & 0.7158                         & 0.5208                         \\ \hline
                               & Hifi-GAN \cite{kong2020hifi}                            & 1.1294                         & 0.9028                         & 0.5744                         \\
\multirow{-2}{*}{Tacotron2 \cite{shen2018natural}}    & Griffin-Lim \cite{griffin1984signal}                        & 1.1916                         & 0.7126                         & 0.5559                         \\ \hline
\multicolumn{2}{c}{VITS \cite{kim2021conditional}}                                             & 0.8788                         & 0.9231                         & 0.4773                         \\ \hline
\end{tabular}
\vspace{-10pt}
\label{tb:vc-intergrity}
\end{center}
\end{table}

\subsection{Generalization} 
\label{sec:robustvc}
We showcase the general effectiveness of our method against different voice cloning attacks described in \Sref{sec:threatmodel-attacker}: professional, regular and low-quality voice cloning attacks. All the speech data of the target speaker are watermarked. To ensure the reliability of our experiments, we randomly generate two different 10-bit watermarks for repeated testing and report the average the results, as shown in \Tref{tb:vc}.

\para{Professional Voice Cloning Attack.}
The attacker trains the acoustic model of voice cloning specifically for the targeted speaker and fine-tunes the vocoder to convert the synthesized mel-spectrograms into the speech that closely resembles the target speaker's voice. 
To simulate such an attack, we adopt Fastspeech2 \cite{ren2020fastspeech} and Tacotron2 \cite{shen2018natural} as the acoustic model, while using Hifi-GAN* \cite{kong2020hifi} fine-tuned on watermarked data (1K segments of 1-10s speech are employed) as the subsequent vocoder. 
For end-to-end single-stage models VITS \cite{kim2021conditional}, we directly train the model using the watermarked dataset.

As shown in the red regions of \Tref{tb:vc}, our method achieves 100\% watermark extraction accuracy across different professional voice cloning attacks. In contrast, \Tref{tb:vc-intergrity} provides the watermark extraction accuracy across the corresponding voice cloning attack on the watermark-free dataset, which behaves like random guessing ($\sim50\%$ ACC). Such comparison highlights the effectiveness of our timbre watermarking solution. In the two tables we also show the quality of synthesized speech, which is calculated between the synthesized speech and the corresponding watermark-free speech of the target speaker. We find that watermarking has negligible impact on the synthesized speech.

\para{Regular Voice Cloning Attack.}
The attacker directly uses a pre-trained vocoder to convert the synthesized mel-spectrogram into the synthesized speech.  Here, we take Hifi-GAN \cite{kong2020hifi} as an example, which is pre-trained on watermark-free speech.
From the blue regions of \Tref{tb:vc}, we observe that our method still achieves a high ACC (over 99\%). In other words, the unwatermarked vocoder Hifi-GAN has less influence on the watermark information, which is still preserved in the subsequent synthesized speech.

\begin{table}[t]
\caption{The impact of different preprocessing operations on the speech quality and robustness of our method.}
\vspace{-0.6em}
\begin{tabular}{cccccc}
\hline
\multirow{2}{*}{Preprocessing}        & \multirow{2}{*}{Parameter}    & \multicolumn{3}{c}{Quality} & \multirow{2}{*}{ACC$\uparrow$} \\ \cline{3-5}
                                   &                           & SNR$\uparrow$      & PESQ$\uparrow$    & SECS$\uparrow$   &                      \\ \hline
\multirow{2}{*}{Resampling}          &\cellcolor{gray!30} 16   kHz &\cellcolor{gray!30} 34.8115  &\cellcolor{gray!30} 4.4967  &\cellcolor{gray!30} 1.0000 &\cellcolor{gray!30} 1.0000               \\
                                   &  8   kHz & 17.1642  & 4.4961  & 0.9025 & 0.9940               \\ \hline
\multirow{4}{*}{Amplitude Scaling} & 20\%                      & 1.9382   & 4.4918  & 0.9575 & 1.0000               \\
                                   & 40\%                      & 4.4368   & 4.4973  & 0.9596 & 1.0000               \\
                                   & 60\%                      & 7.9589   & 4.4986  & 0.9772 & 1.0000               \\
                                   & 80\%                      & 13.9790  & 4.4991  & 0.9942 & 1.0000               \\ \hline
\multirow{8}{*}{MP3 Compression}   &\cellcolor{red!30} 8 kbps                    &\cellcolor{red!30} 9.0414   &\cellcolor{red!30} 2.2115  &\cellcolor{red!30} 0.7565 &\cellcolor{red!30} 0.9186               \\
                                   & 16 kbps                   & 13.1554  & 3.3484  & 0.9552 & 0.9992               \\
                                   & 24 kbps                   & 15.2631  & 3.9259  & 0.9888 & 0.9999               \\
                                   & 32 kbps                   & 17.2272  & 4.0695  & 0.9962 & 1.0000               \\
                                   & 40 kbps                   & 18.7795  & 4.1902  & 0.9975 & 1.0000               \\
                                   & 48 kbps                   & 20.8746  & 4.3122  & 0.9986 & 1.0000               \\
                                   & 56 kbps                   & 22.8885  & 4.3813  & 0.9991 & 1.0000               \\
                                   &\cellcolor{gray!30} 64 kbps                   &\cellcolor{gray!30} 23.9958  &\cellcolor{gray!30} 4.4136  &\cellcolor{gray!30} 0.9992 &\cellcolor{gray!30} 1.0000               \\ \hline
Recount                            & 8   bps                   & 22.9103  & 3.1708  & 0.9757 & 0.9995               \\ \hline
\multirow{4}{*}{Median Filtering}  & 5   Samples               & 14.8666  & 3.6664  & 0.9459 & 1.0000               \\
                                   & 15 Samples                & 8.9079   & 2.5726  & 0.7875 & 0.9933               \\
                                   & 25 Samples                & 5.3999   & 2.1427  & 0.7338 & 0.9806               \\
                                   & 35 Samples                & 3.2550   & 1.8721  & 0.6861 & 0.9402               \\ \hline
Low Pass Filtering                 &\cellcolor{red!30} 2000 Hz                   &\cellcolor{red!30} 12.8558  &\cellcolor{red!30} 3.8824  &\cellcolor{red!30} 0.7280 &\cellcolor{red!30} 0.9030               \\ \hline
High Pass Filtering                & 500   Hz                   & 3.7635   & 3.7919  & 0.6551 & 1.0000               \\ \hline
\multirow{5}{*}{Gaussian Noise}    
                                   & 20 dB                     & 20.0002  & 3.1287  & 0.9104 & 0.9962               \\
                                   & 25 dB                     & 24.9989  & 3.5182  & 0.9670 & 0.9995               \\
                                   & 30 dB                     & 29.9981  & 3.8662  & 0.9919 & 1.0000               \\
                                   & 35 dB                     & 34.9941  & 4.1277  & 0.9981 & 1.0000               \\
                                   & 40 dB                     & 39.9888  & 4.3038  & 0.9994 & 1.0000               \\ \hline
\end{tabular}
\label{tb:common-distortion}
\end{table}

\para{Low-quality Voice Cloning Attack.}
The attacker trains an acoustic model on the target speaker's speech data to synthesize mel-spectrograms, which are then transferred into speech waveforms using the Griffin-Lim vocoder. 
In this case, the mel-spectrograms synthesized by the acoustic model contain the watermark information, and the Griffin-Lim algorithm does not include any trainable parameters. Therefore, it does not tend to transfer watermarked mel-spectrograms to non-watermarked waveforms, as the watermark-free pre-trained vocoder Hifi-GAN does. 
As shown in the grey regions of \Tref{tb:vc}, the watermark extraction ACC can still reach 100\%, which is even slightly better than the regular voice cloning attack.

\begin{figure}[t] 		
    \centering	
    \includegraphics[scale=0.3]{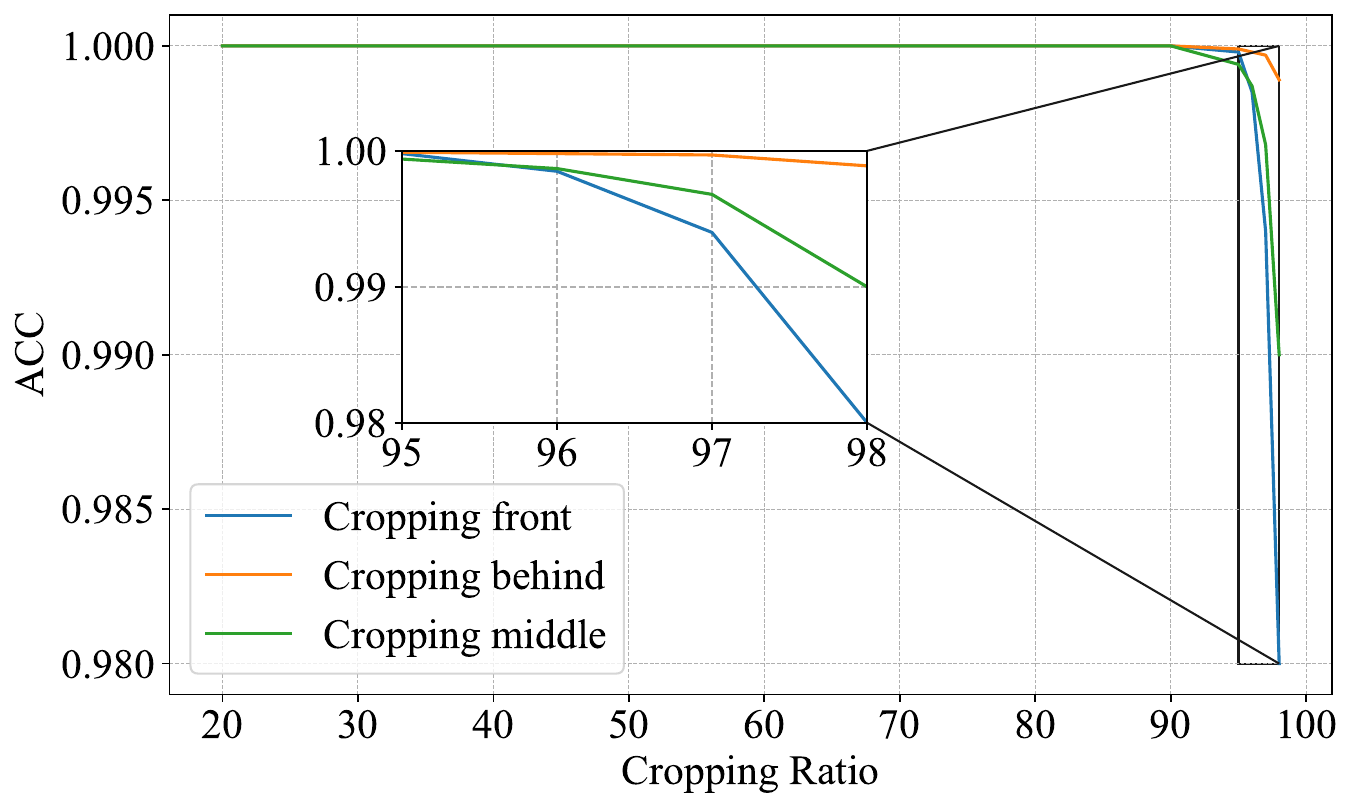}
    \vspace{-0.6em}
    \caption{Robustness against different cropping strategies.}
    \label{fig:crop-robustness}
    \vspace{-15pt}  
\end{figure}

\subsection{Robustness}
\label{sec:common}
We evaluate the robustness of our approach against unseen 
distortions including cropping and other preprocessing operations.
Robustness against cropping is particularly important because it is an unavoidable and intensive preprocessing operation in data collection before voice cloning attacks.
\Fref{fig:crop-robustness} shows the ACC curve with different cropping ratio, when the watermarked audio is cropped from the front, middle and behind.
\hl{
\Fref{fig:crop-example} in the Appendix shows an example of cropping audio. 
}
We observe that even when 90\% of the audio is cropped, the watermark can still be extracted with 100\% accuracy.
It is worth mentioning that different from the synchronization-preserving cropping used in \cite{su2018snr} and \cite{liu2022dear}, we directly crop off the audio to make it lose synchronization in the time dimension. Thus, the above results also indicate that our watermarking scheme can resist desynchronization attacks as it is intrinsically robust to distortions in the time domain. 
We further test whether our scheme can withstand other common audio processing distortions.
As shown in \Tref{tb:common-distortion}, the audio watermarked by our method can resist various preprocessing operations, and we obtain above 90\% ACC in all cases.

\begin{table}[t]
\begin{center}
\caption{Robustness against voice cloning attacks with regular and harmful preprocessing.}
\vspace{-0.6em}
\begin{tabular}{ccccc}
\hline
\multicolumn{2}{c}{Pre-processing}        & PESQ$\uparrow$   & SECS$\uparrow$   & ACC$\uparrow$    \\ \hline
\multirow{3}{*}{Regular} & Resampling 16K          & 1.0775 & 0.9122 & 1.0000 \\
& Mp3 Compression   64kbps & 1.0347 & 0.9077 & 1.0000 \\
& Combined              & 1.0776 & 0.9064 & 1.0000 \\ \hline
\multirow{3}{*}{Harmful} & Mp3 Compression   8kbps      & 0.8284 & 0.6675 & 0.8996 \\ 
& Low Pass   Filtering 2000 Hz & 1.0836 & 0.6481 & 0.9482 \\
& Combined                  & 1.0324 & 0.6567 & 0.9144 \\ \hline
\end{tabular}
\label{tb:regular}
\end{center}
\vspace{-12pt}
\end{table}

\subsection{Resistance Against Adaptive Attacks} \label{sec:adaptive}
\label{sec:experiment-adaptive}
We consider different adaptive attacks in \Sref{sec:adp}, and validate the corresponding resistance of our solution. 

\noindent\textbf{Preprocessing Before Voice Cloning Attack.}
In this scenario, the attacker wants to apply some preprocessing operations on the training audio before voice cloning attacks.
According to the influence on the audio quality, we further categorize the attacks into two parts: regular preprocessing that does not lead to significant quality degradation, and harmful preprocessing that has a severe impact on the audio quality. For example, in \Tref{tb:common-distortion}, Resampling 16K and MP3 compression 64k (highlighted in gray) are regular processing operations, while MP3 compression 8k 
and Low-pass Filtering 2000 Hz (highlighted in red) belong to harmful preprocessing operations.

We further consider two cases: a) the attacker adopts some common regular preprocessing operations before voice cloning, trying to occasionally bypass the watermark detection; b) the attacker leverages harmful preprocessing to intentionally degrade watermarked data before voice cloning. We present the corresponding results for the above cases in  \Tref{tb:regular} (VITS \cite{kim2021conditional} is utilized as the default voice cloning attack model). It is noted that ``Combined'' signifies the application of a randomly selected preprocessing approach to each individual audio data sample.
In particular, regular preprocessing does not influence the quality of the synthesized speech (above 0.90 SECS), and our watermark extraction accuracy reaches 100\% for all. For harmful preprocessing, our proposed method can still achieve nearly 90\% ACC, even though the quality of the synthesized speech is destroyed significantly (nearly 0.65 SECS). 
Some synthesized samples can be found on our project website.

\begin{figure}[t] 		
    \centering	
    \includegraphics[scale=0.3]{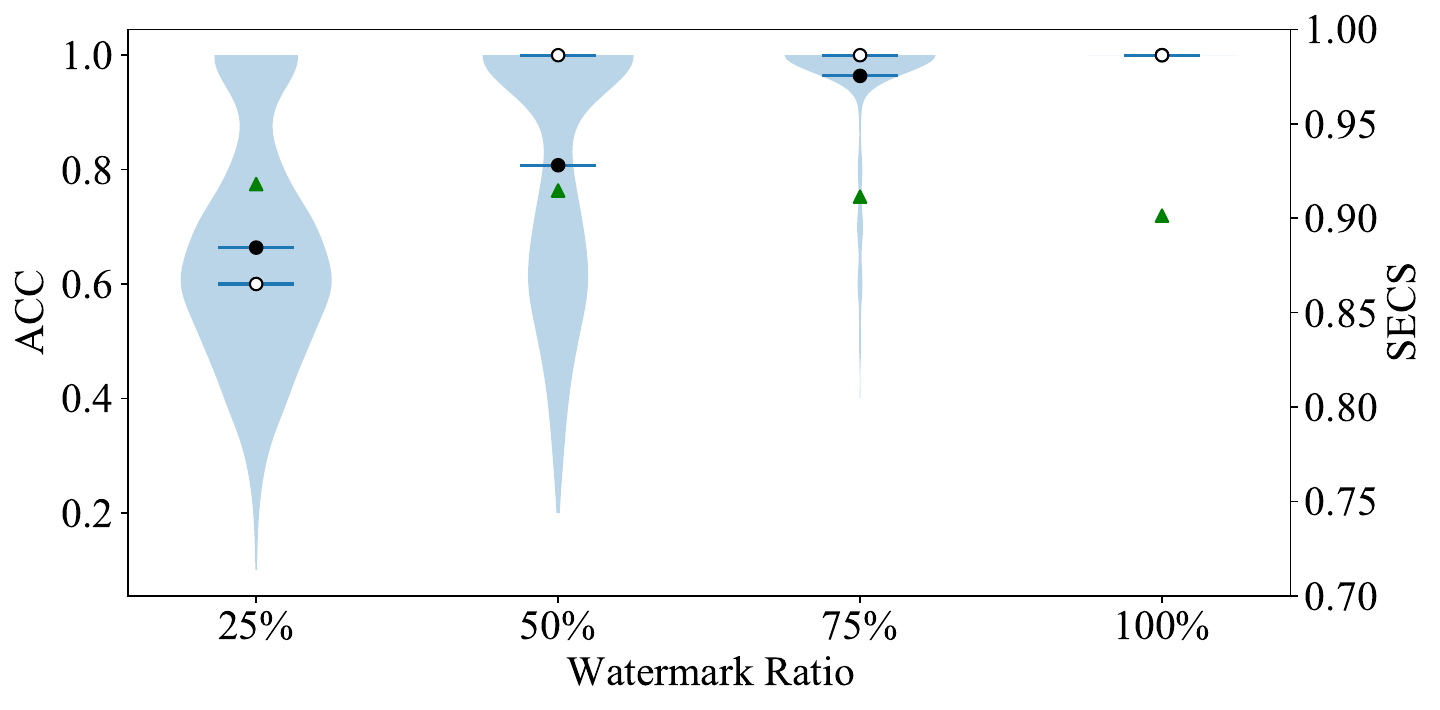}
    \vspace{-2mm}
    \caption{The performance of our method against voice cloning attacks with different watermark ratios.
    Black dots and white dots indicate mean accuracy and median accuracy, respectively. Green triangles represent the average SECS values of synthesized speech.}
    \vspace{-15pt}
    \label{fig:wm-ratio}
\end{figure}

\begin{figure*}[t]
    \hspace{0.3cm}
    \begin{tabular}[b]{c@{}c@{}}
    \includegraphics[width=0.28\textwidth, scale=0.9]{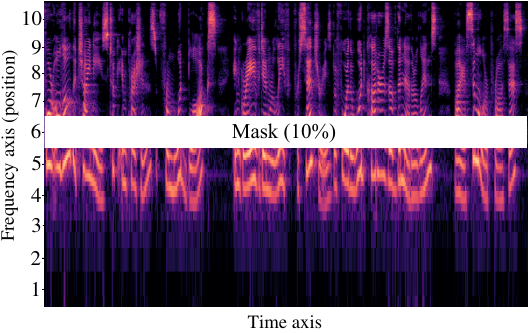}\label{fig:sub1} \\[-0.05ex]
    \hspace{0.7cm}(a)
    \end{tabular}
    \hspace{0.7cm}
    \begin{tabular}[b]{c@{}c@{}}
    \includegraphics[width=0.28\textwidth,scale=0.30]{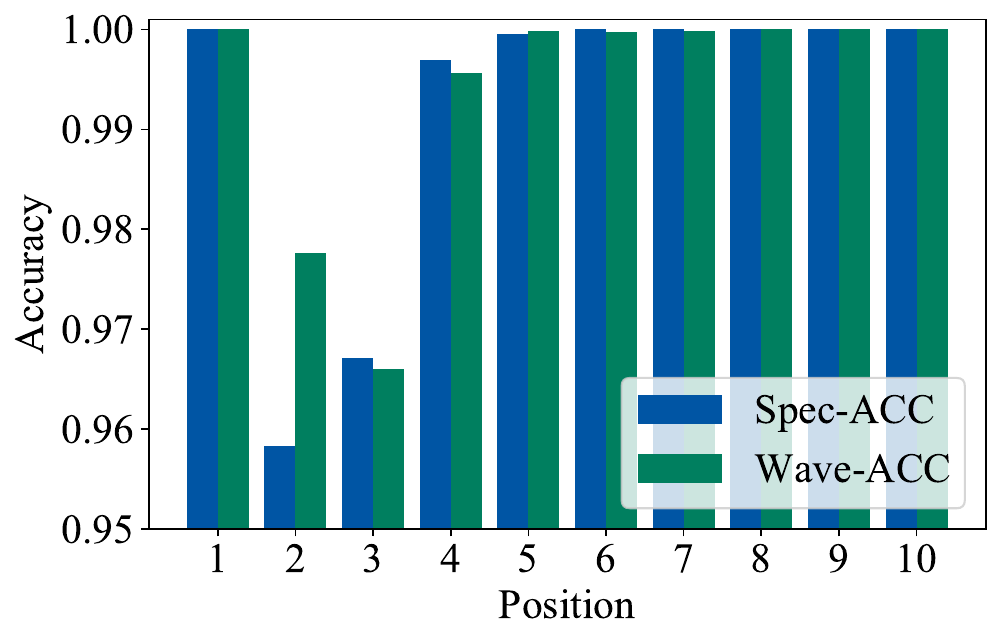}\label{fig:sub2} \\[-0.05ex]
    \hspace{0.9cm}(b)
    \end{tabular}
    \hspace{0.7cm}
    \begin{tabular}[b]{c@{}c@{}}
    \includegraphics[width=0.28\textwidth,scale=0.4]{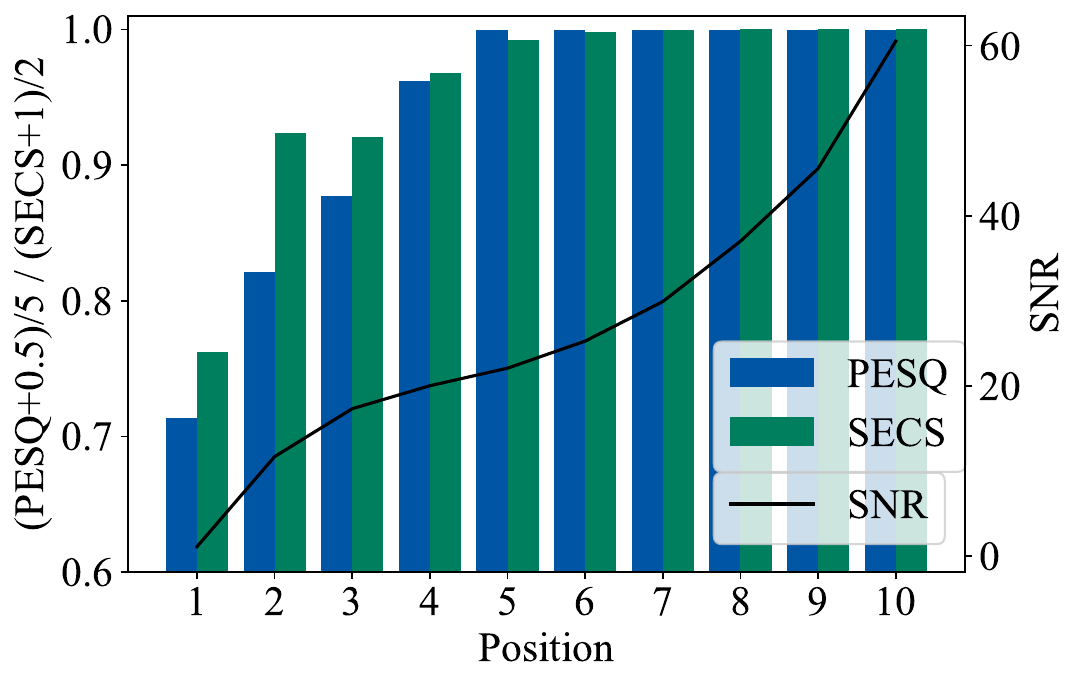}\label{fig:sub3} \\[-0.05ex]
    (c)
    \end{tabular}
        \vspace{-2mm}   
    \caption{(a) Visual example of the spectrogram with 10\% masked. (b) Watermark extraction accuracy when different frequency bands of the spectrogram are masked. (c) Speech quality when different frequency bands of the spectrogram are masked.}
    \vspace{-5pt}  
    \label{fig:three_images}
\end{figure*}

\begin{figure}[t] 		
    \centering	
    \includegraphics[scale=0.3]{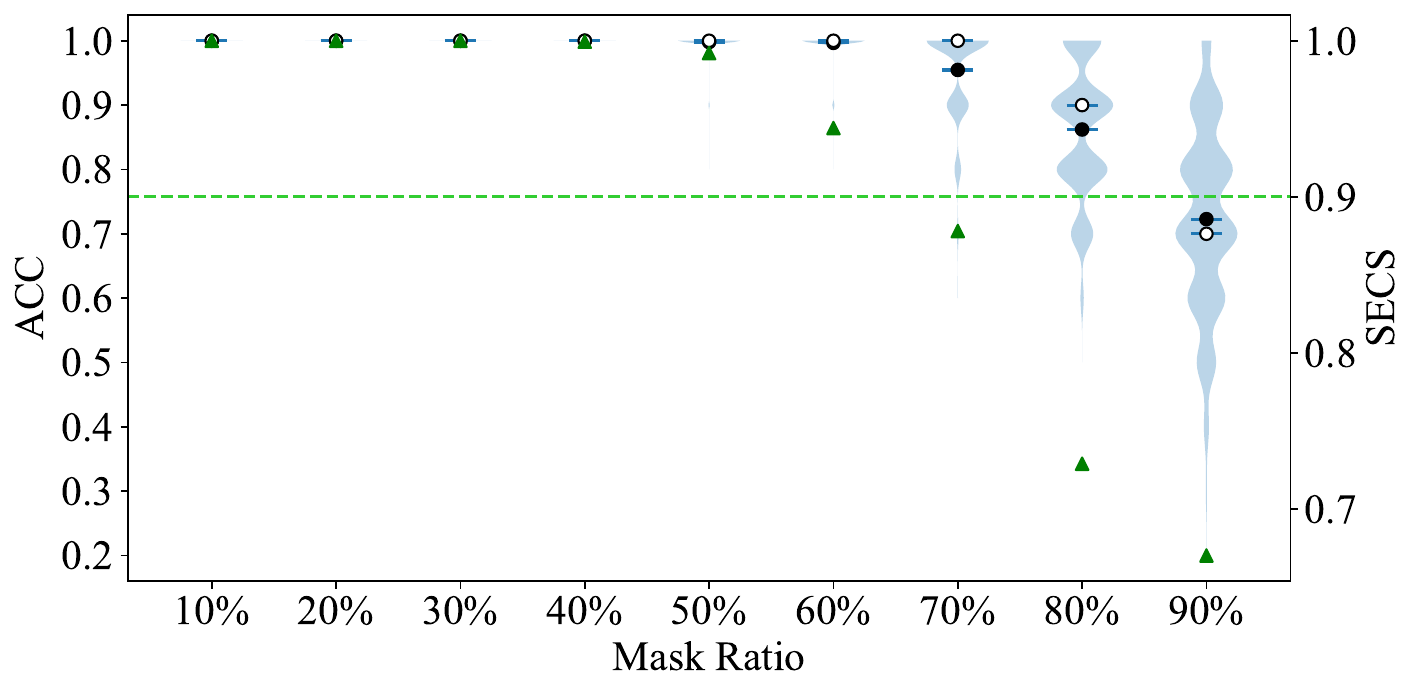}
    \vspace{-0.8em}
    \caption{The performance of our method against different frequency masking ratios. Black dots and white dots indicate mean ACC and median ACC, respectively. The green triangles represents the average SECS values of synthesized speech, while the green line indicates the bypass authentication line.}
    \label{fig:mask-ratio}
        \vspace{-15pt}
\end{figure}

\noindent\textbf{Voice Cloning Attack with Partial Unwatermarked Data.}
The attacker may have access to some unwatermarked speech data of the target speaker.
To explore the effectiveness of our approach in such a scenario, we train speech synthesis models on datasets with different watermark ratios. To expedite the experimental process, we use the end-to-end speech synthesis model, VITS \cite{kim2021conditional}, as the default attack model.
\Fref{fig:wm-ratio} shows the corresponding watermark extraction accuracy. We observe that with a higher ratio of watermarked speech dataset, the synthesized speech contains more complete watermark information. A 75\% ratio of watermarked speech data is sufficient to achieve an extraction accuracy of over 95\% (96.38\%) in synthesized speech, while only a 25 \% ratio of watermarked speech data can still retain 66.36\% of watermark information, which is still effective in detecting such adaptive attack.

\begin{table}[t]
\begin{center}
\caption{Robustness against watermark overwriting attacks.  \# indicates that the attacker extracts the watermark by his own extractor. * indicates that the attacker trains his own embedding and extraction models guided by the proposed method.}
\vspace{-0.6em}
\begin{tabular}{ccc@{\extracolsep{\fill}}>{\centering\arraybackslash}p{1.5cm}@{\extracolsep{\fill}}>{\centering\arraybackslash}p{1.5cm}@{\extracolsep{\fill}}>{\centering\arraybackslash}p{1.5cm}}
\hline
Method                                      & PESQ$\uparrow$   & SECS$\uparrow$   & wm1-ACC$\uparrow$       & wm2-ACC$\downarrow$      & wm2-ACC\#$\downarrow$     \\ \hline
FSVC \cite{zhao2021desynchronization}       & 1.0334 & 0.9115 & 1.0000 ($\checkmark$)   & 0.4000 ($\times$)     & 0.5544 ($\times$)     \\
RFDLM \cite{liu2018patchwork}               & 1.0727 & 0.9102 & 1.0000 ($\checkmark$)   & 0.4000 ($\times$)     & 0.4986 ($\times$)     \\ 
The Proposed                                & 0.9891 & 0.8968 & 0.4000 ($\times$)       & 1.0000 ($\checkmark$) & 1.0000 ($\checkmark$) \\
The Proposed*                               & 0.9951 & 0.8789 & 0.9346 ($\checkmark$)   & 0.4646 ($\times$)     & 1.0000 ($\checkmark$) \\
\hline
\end{tabular}
\label{tb:rewm}
\end{center}
\vspace{-15pt}
\end{table}

\noindent\textbf{Watermark Overwriting Attack.}
With the collected watermarked audio data, the attacker can further embed his own watermark before conducting the voice cloning attack. 
In our threat model, voice cloning attacks are detected by the platform's extractor. Nevertheless, we also show the results based on the attacker's extractor. 
Specifically, we embed the original watermark (wm1) using our proposed method. To further embed the attacker's watermark (wm2), we first adopt existing audio watermarking schemes FSVC \cite{zhao2021desynchronization} and RFDLM \cite{liu2018patchwork}. As shown in \Tref{tb:rewm}, the original watermark can still be extracted as the evidence of voice cloning attacks, while the attacker fails to extract his watermark. This is because FSVC \cite{zhao2021desynchronization} and RFDLM \cite{liu2018patchwork} are fragile to voice cloning attacks, which do not interrupt our watermark signal during the attacks.

In addition, we further assume that the attacker can utilize our proposed method for watermark overwriting. If the attacker is an insider malicious user, he can directly utilize the well-trained embedding and extraction models by our method.
Otherwise, he can train his own model with the proposed strategy, which is denoted as ``the proposed *" in \Tref{tb:rewm}. The results show that our method will fail to resist the overwriting attack only when an inner attacker exists. 
\hl{
To address this issue, we further design a weighted embedding process, which is controlled by the subsequent watermark decoder, and add the watermark overwriting distortion in the original distortion layer to fine-tune the model.
As shown in \Tref{tb:exp-rewm},
the new strategy will not degrade the audio quality and achieve a 100\% ACC against overwriting attacks.
More details are in \ASref{sec:appendix-overwriting}.
}

\para{Audio Reconstruction-based Removal Attacks.} 
\hl{
We try to compromise the watermark by audio reconstruction, \ie, the adaptive attacker may try to train a reconstruction model with many watermarked-unwatermarked data pairs. For the audio reconstruction model, we adopt the state-of-the-art popular MelVAE\footnote{\href{https://github.com/moiseshorta/MelSpecVAE}{https://github.com/moiseshorta/MelSpecVAE}}. For data pairs, we use 10, 000 audios from Librispeech \cite{panayotov2015librispeech} as non-watermarked audio, and embed random watermarks into them by the proposed method to obtain the corresponding watermarked audio. Then, we train MelVAE with such data pairs in a supervised manner, aiming to transform watermarked audio into non-watermarked one.
We observe that such an audio reconstruction model can only produce speech with poor quality (some samples are shown on our web page). Nevertheless,  the accuracy of watermark extraction can still reach 74.90\%.}

\hl{
Besides, we consider an existing audio watermark removal attack using a Variational Autoencoder (VAE) pre-trained on a clean audio dataset \cite{zhao2023generative} and test our method under such an attack. Specifically,  we use the VAE of the audio generation model \cite{liu2023audioldm} for the reconstruction operation. Note that, as described by the authors, this VAE was trained on AudioSet~(AS)~\cite{gemmeke2017audio}, AudioCaps~(AC)~\cite{kim2019audiocaps}, Freesound~(FS)\footnote{\url{https://freesound.org/}}, and BBC Sound Effect library~(SFX)\footnote{\url{https://sound-effects.bbcrewind.co.uk/search}}, and exhibits good audio generation quality.
Based on the experimental results in \Fref{fig:tango-vae}, we can see that while VAE reconstruction indeed has some effect on the watermark, it is almost unable to erase it (Average ACC:100\% $\rightarrow$ 99.98\%). \Fref{fig:tango-vae-spectrum} shows two examples of the effect of VAE reconstruction on the audio signal. The reconstructed speech samples and the samples cloned based on them are accessible on our web page.}

\noindent\textbf{Identifying Watermark Location and then Removing It.}
As illustrated in \Fref{fig:three_images}-a, we apply a 10\% frequency band masking operation to different frequency locations (from bottom to top) of the watermarked speech spectrogram.

In \Fref{fig:three_images}-b, we present the watermark extraction accuracy obtained directly from both the masked spectrogram (Spec-ACC) and after re-extracting the spectrogram from the ISTFT-transformed audio signal (Wave-ACC). Based on experimental results, we find that masking at the medium-to-low frequency range (\ie, Position 2 and 3) has the most significant impact on the watermark extraction ability, and conjecture the watermark is primarily embedded in the medium-to-low frequency range. 

Therefore, we conduct an adaptive attack: masking such frequency range as preprocessing before voice cloning, and then adopting VITS \cite{kim2021conditional}  to launch the attack. The results prove that this adaptive attack still cannot defeat our method: watermark extraction accuracy from synthesized speeches is $100\%$ (masking position2) and $92.14\%$ (masking position3).
\Fref{fig:three_images}-c shows the impact of masking at different locations on the audio quality (SNR, PESQ, SECS). We find that masking at lower frequencies has a more severe effect on the audio quality, and masking at Positions 2 and 3 has a particularly significant impact. As the masked frequency range increases, the audio quality improves.
To remove watermarks strongly, we try masking more frequency bands of the spectrogram (from high to low frequencies). We gradually increase the percentage of the mask from the top of the spectrogram, since masking low frequencies introduces a very serious quality loss. As shown in \Fref{fig:mask-ratio}, as the mask ratio increases from 10\% to 90\%, although the ACC degrades from 100\% to near 70\%, the audio quality is also destroyed by a large margin (SECS from 1.0 to 0.1).
\hl{For a more comprehensive evaluation, we further adopt 5\% and 20\% masking ratios. The corresponding results are shown in \Fref{fig:mask}  of \ASref{sec:mask-ratio-2}, which are consistent with the results with a 10\% masking ratio.}

\noindent\textbf{Voice Cloning Attack with Public Extractor.}
Because the extractor is publicly released by the platform, the attacker can attempt to generate audios to bypass it.
In practice, the extractor may be packaged as an API, and the attacker only knows whether the generated audio deceives the extractor. To simulate this scenario, the attacker can first train a binary classifier to distinguish the watermarked and unwatermarked data. With the well-trained classifier, he is able to train his TTS model constrained by a domain adversarial loss, which forces the model to output unwatermarked data. We find such a strategy (\ie, domain-adversarial training) will degrade the performance of the TTS model (SECS from 0.9014 to 0.8844), while our method can still achieve 100\% extraction accuracy.

\noindent\textbf{Combing Multiple Attack Strategies.}  \label{sec:strategy}
We further consider combining different attack strategies to destroy the proposed method.  This entails the integration of diverse attack schemes, encompassing regular preprocessing, harmful preprocessing, domain-adversarial training, VAE reconstruction and watermark overwriting. In a ntushell, more severe attack strategies will further destroy the utility of voice cloning, while the proposed method is still somewhat effective. For example, taking resampling 16 kHz as pre-processing and MP3 compression 16 kbps as post-processing, compared with only pre-processing, ACC suffers a slight degradation (ACC:100\% $\rightarrow$ 99.94\%) but the quality degrades by a large margin (SECS: 1.000 $\rightarrow$ 0.8575). The specific experimental results can be found on the paper's website.

\begin{table*}[t]
\begin{center}
\centering
\caption{The performance of our proposed method against voice cloning attacks on commercial platforms.}
            \vspace{-1em}
\begin{tabular}{c|c|ccccccc}
\hline
Service & Language & Metric & \multicolumn{6}{c}{Speaker} \\ \hline
                                  &                           &  & P225                          & P226                          & P227                          & P228                          & P229                          & P230                          \\ \cline{4-9} 
                                  &                           & PESQ$\uparrow$    & 2.5958                        & 2.7235                        & 2.3573                        & 2.3235                        & 2.7419                        & 1.7095                        \\
                                  &                           & SECS$\uparrow$    & 0.8611                        & 0.8701                        & 0.8552                        & 0.8537                        & 0.8592                        & 0.8519                        \\
                                  & \multirow{-4}{*}{English} & ACC$\uparrow$     & 1.0000  & 1.0000  & 1.0000  & 1.0000  & 1.0000  & 1.0000  \\ \cline{2-9} 
                                  &                           & & D4                            & D6                            & D7                            & D8                            & D11                           & D12                           \\ \cline{4-9} 
                                  &                           & PESQ$\uparrow$    & 1.7642                        & 1.9851                        & 2.6490                        & 2.0223                        & 2.3808                        & 1.2313                        \\
                                  &                           & SECS$\uparrow$    & 0.7836                        & 0.8034                        & 0.7622                        & 0.8219                        & 0.7304                        & 0.7103                        \\
\multirow{-8}{*}{PaddleSpeech \cite{zhang2022paddlespeech} }    & \multirow{-4}{*}{Chinese} & ACC$\uparrow$     & 1.0000  & 1.0000  & 1.0000  & 1.0000  & 1.0000  & 1.0000  \\ \hline
                                  &                           & & P225                          & P226                          & P227                          & P228                          & P229                          & P230                          \\ \cline{4-9} 
                                  &                           & PESQ$\uparrow$    & 0.7809                        & 1.5610                        & 1.1913                        & 1.1684                        & 1.2601                        & 1.2694                        \\
                                  &                           & SECS$\uparrow$    & 0.7576                        & 0.8564                        & 0.7324                        & 0.8781                        & 0.8495                        & 0.8799                        \\
\multirow{-4}{*}{Voice-Cloning-App \cite{vca}} & \multirow{-4}{*}{English} & ACC$\uparrow$     & 0.9000 & 0.9100 &  0.9000 & 0.9000 & 0.9500 & 0.9200 \\ \hline
\end{tabular}
\label{tb:api}
\end{center}
\vspace{-15pt}
\end{table*}

\subsection{Practicality in Real-world Services} \label{sec:vc-tools}
In practice, the attacker may directly leverage real-world services to conduct voice cloning attacks in an efficient and convenient way. Specifically, he can collect a small number of the target speaker's voice samples and upload them to real-world services to implement customized voice cloning based on their instructions. When the collected voice samples have been watermarked by our method, we are able to extract the watermark information from the synthetic voice. We test both real-world TTS service and voice conversion service.

\begin{table*}[t]
\begin{center}
\caption{Robustness comparison between Distortion-Blind Watermarking Model (DBWM) and Full Model against different voice cloning attacks. The red, blue, and gray areas represent professional, regular, and low-quality voice cloning attacks, respectively. The quality of synthesized speech is also provided.}
\vspace{-0.6em}
\begin{tabular}{c|c|cc | cc | c}
\hline
                             &                          & \multicolumn{2}{c|}{Fastspeech2* \cite{ren2020fastspeech}}          & \multicolumn{2}{c|}{Tacotron2* \cite{shen2018natural}}            & \cellcolor[HTML]{FFCCC9}      \\ \cline{3-6}
\multirow{-2}{*}{Model}      & \multirow{-2}{*}{Metric} & \cellcolor[HTML]{DAE8FC}{Hifi-GAN \cite{kong2020hifi}} & \cellcolor[HTML]{EFEFEF}{Griffin-Lim}  \cite{griffin1984signal}         & \cellcolor[HTML]{DAE8FC}{Hifi-GAN \cite{kong2020hifi}} & \cellcolor[HTML]{EFEFEF}{Griffin-Lim} \cite{griffin1984signal}     & \cellcolor[HTML]{FFCCC9} \multirow{-2}{*} {VITS* \cite{kim2021conditional}} \\ \hline
                             & PESQ$\uparrow$   & \cellcolor[HTML]{DAE8FC}{1.0238}     & \cellcolor[HTML]{EFEFEF}{1.0765}      & \cellcolor[HTML]{DAE8FC}{1.2018}   & \cellcolor[HTML]{EFEFEF}{1.3184}    & \cellcolor[HTML]{FFCCC9}{1.0299}     \\
                             & SECS$\uparrow$   & \cellcolor[HTML]{DAE8FC}{0.8996}     & \cellcolor[HTML]{EFEFEF}{0.7152}      & \cellcolor[HTML]{DAE8FC}{0.8858}   & \cellcolor[HTML]{EFEFEF}{0.7025}    & \cellcolor[HTML]{FFCCC9}{0.9173}     \\
\multirow{-3}{*}{DBWM}       & ACC$\uparrow$    & \cellcolor[HTML]{DAE8FC}{0.6508}     & \cellcolor[HTML]{EFEFEF}{0.6136}      & \cellcolor[HTML]{DAE8FC}{0.6854}   & \cellcolor[HTML]{EFEFEF}{0.6108}    & \cellcolor[HTML]{FFCCC9}{1.0000}     \\ \hline
                             & PESQ$\uparrow$   & \cellcolor[HTML]{DAE8FC}{1.0770}     & \cellcolor[HTML]{EFEFEF}{1.1224}      & \cellcolor[HTML]{DAE8FC}{1.1350}   & \cellcolor[HTML]{EFEFEF}{1.2342}    & \cellcolor[HTML]{FFCCC9}{1.0561}     \\
                             & SECS$\uparrow$   & \cellcolor[HTML]{DAE8FC}{0.8958}     & \cellcolor[HTML]{EFEFEF}{0.7075}      & \cellcolor[HTML]{DAE8FC}{0.8440}   & \cellcolor[HTML]{EFEFEF}{0.7083}    & \cellcolor[HTML]{FFCCC9}{0.9014}     \\
\multirow{-3}{*}{Full Model} & ACC$\uparrow$    & \cellcolor[HTML]{DAE8FC}{0.9978}     & \cellcolor[HTML]{EFEFEF}{1.0000}      & \cellcolor[HTML]{DAE8FC}{0.9998}   & \cellcolor[HTML]{EFEFEF}{1.0000}    & \cellcolor[HTML]{FFCCC9}{1.0000}     \\ \hline
\end{tabular}
\label{tb:ab}
\end{center}
\vspace{-15pt}
\end{table*}

\para{Real-world TTS Service.}
We employ two widely-used speech synthesis tools, PaddleSpeech \cite{zhang2022paddlespeech} and Voice-Cloning-App \cite{vca}, to conduct voice cloning attacks in a black-box way. 
For PaddleSpeech, we use the tool through the Baidu Paddle AI Studio platform \cite{PaddleSpeech}. The attacker can easily fine-tune the speech synthesis model and customize the cloned voice of the target speaker with just a few clicks.
For Voice-Cloning-App \cite{vca}, we use the released executable software provided by the author, following the software's documentation for voice cloning.
For both services, the attacker only needs 10 segments of the target speaker's speech within 10 seconds. Additionally, we follow the work \cite{wenger2021hello} by using 10 segments of texts as input for the synthesized speech (listed in \Tref{tb:phrases} and \Tref{tb:phrases-cns} in the Appendix). We extract watermarks from the synthesized speech and verify the extraction accuracy as well as the quality of the synthesized speech. For comprehensive evaluation, we test multiple speakers (p225 $\sim$  p230, the first six speakers) from the VCTK dataset \cite{veaux2017cstr}. To explore the efficacy of our approach in different languages, we also select the first six speakers (D4, D6, D7, D8, D11, D12) from the THCHS30 \cite{wang2015thchs} test set and test the more challenging Chinese voice cloning scenario using the PaddleSpeech tool, an open source tool that additionally supports Chinese besides English.

\Tref{tb:api} presents the evaluation results. Specifically, 
for PaddleSpeech \cite{zhang2022paddlespeech}, the quality of the synthesized voice is considerably superior. We can achieve an exceptional watermark extraction accuracy ($ACC=100\%$).
For Voice-Cloning-App \cite{vca}, although the quality of the synthesized voice is inferior, the watermark extraction accuracy is still maintained at a satisfactory level ($ACC>=90\%$), which further substantiates the outstanding effectiveness of the proposed watermarking methodology. The synthesized speech samples from these services can be found on our project website.

\para{Real-world Voice Conversion Service.} \label{sec:vc}
Very recently, the voice conversion service 
so-vits-svc \cite{svs}
becomes increasingly popular for singing song synthesis. Thus, we also try our method in such a scenario and validate its effectiveness.
Specifically, we use the first $30$ singing voices (around $5s$ per voice) from the Opencpop dataset \cite{wang2022opencpop}, embed watermarks in them, and then use these voices to train so-vits-svc according to the instruction document. After that, we use the trained model to perform voice conversion on a song. Here, we use the original song ``Right Here Waiting" from Richard Marx \cite{jz},
and use Ultimate Vocal Remover \cite{uvr}
to remove the background music. Then, we divide it into $24$ segments ($10s$ per segment) and perform the voice conversion for each segment. The vocal data are available on our website. The final extraction accuracy of each segment of the synthesized vocals is $100\%$, which further proves the effectiveness of our method in real speech cloning scenarios.

\begin{figure}[t] 		
    \centering	
    \includegraphics[scale=0.3]{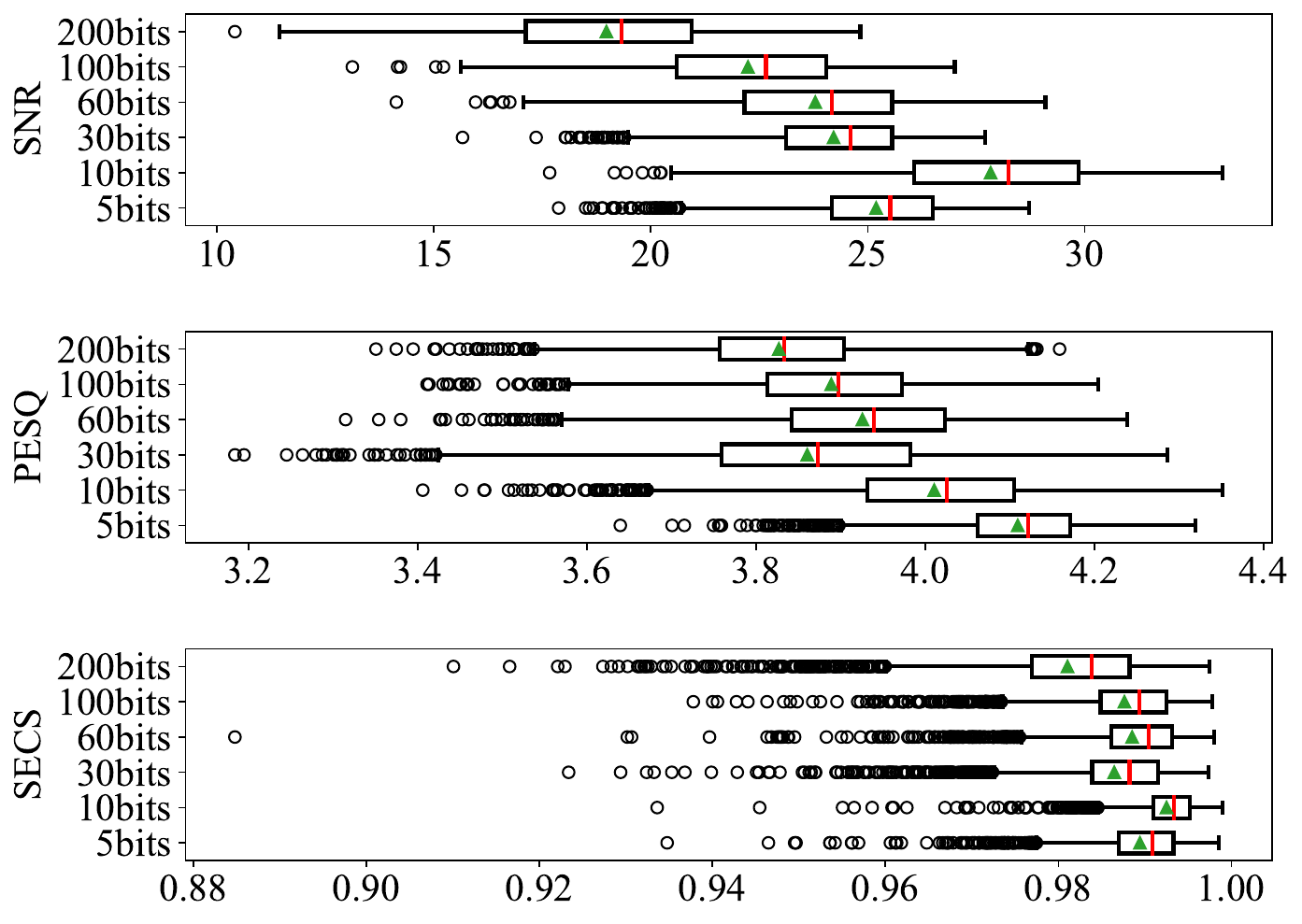}
    \vspace{-0.8em}
    \caption{The fidelity of our method with different embedded watermark bits. Green triangles represent the mean values and red lines indicate the median values.}
    \vspace{-14pt}
    \label{fig:emratio-fidelity}
\end{figure}

\begin{figure}[t] 		
    \centering	
    \includegraphics[scale=0.3]{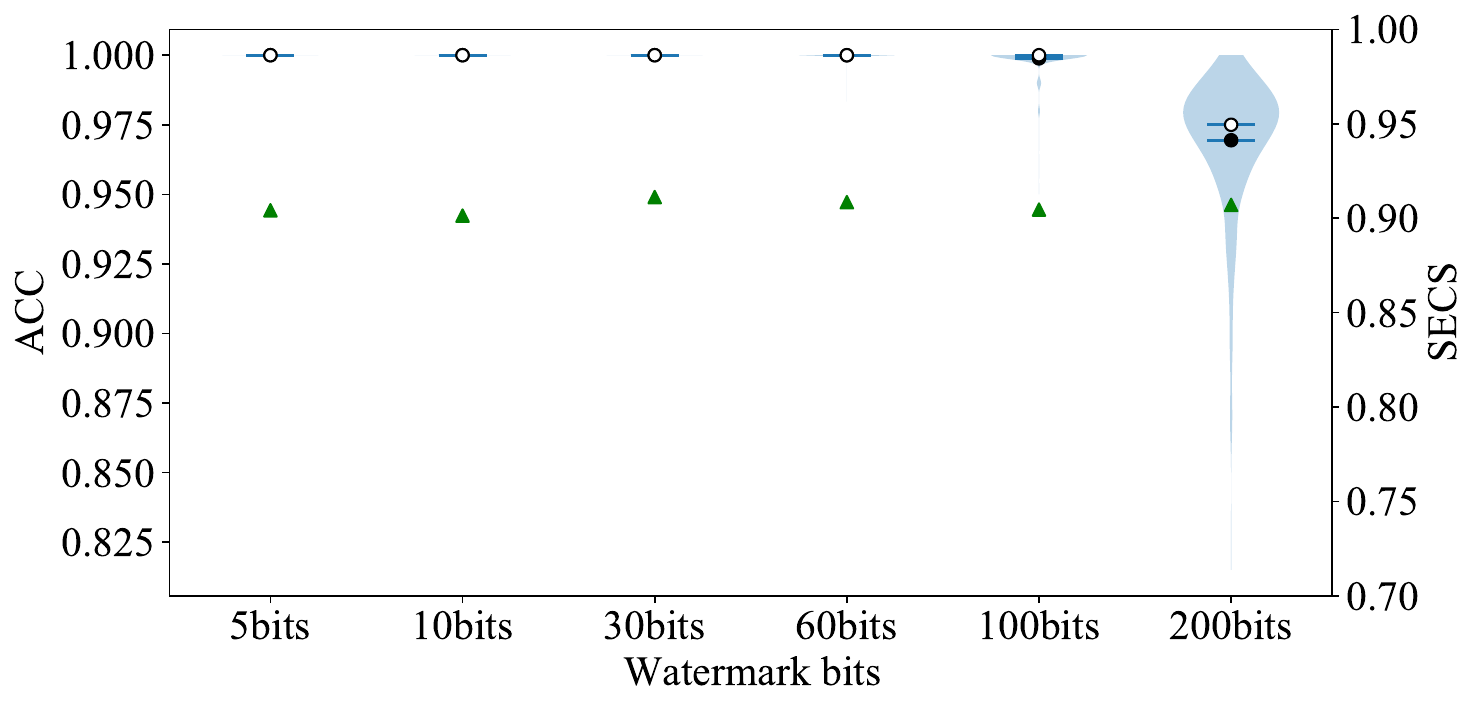}
                \vspace{-0.8em}
    \caption{Watermark extraction accuracy of our method with different embedded watermark bits. 
        Black dots and white dots indicate mean accuracy and median accuracy, respectively, while green triangles represent the average SECS values of synthesized speech.}
    \vspace{-16pt}
    \label{fig:emratio-acc}
\end{figure}

\subsection{Ablation Study}
\label{sec:ablation-study}

\para{Influence of Watermark Bits.}
To investigate the impact of the watermark length, 
we attempt to embed bit strings of varying lengths (5 bits, 10 bits, 30 bits, 60 bits, 100 bits, 200 bits) in the carrier audio and utilize VITS \cite{kim2021conditional} as the default voice cloning attack model. 
\Fref{fig:emratio-fidelity} shows the impact of different watermark bits on the quality of the watermarked speech. As we can see, embedding more bits leads to a decrease in fidelity, but this trend is not significant, especially for the impact on timbre (SECS). This means that we can flexibly choose the embedding capacity for different scenarios.
\Fref{fig:emratio-acc} further shows the watermark extraction accuracy ACC under the voice cloning attacks. As the number of embedded bits increases, ACC still keeps near 100\%, except when the embedded watermark bits is extremely large, \ie, 200 bits, where ACC is also higher than 95\%.

\begin{figure}[] 		
    \centering	
    \includegraphics[scale=0.3]{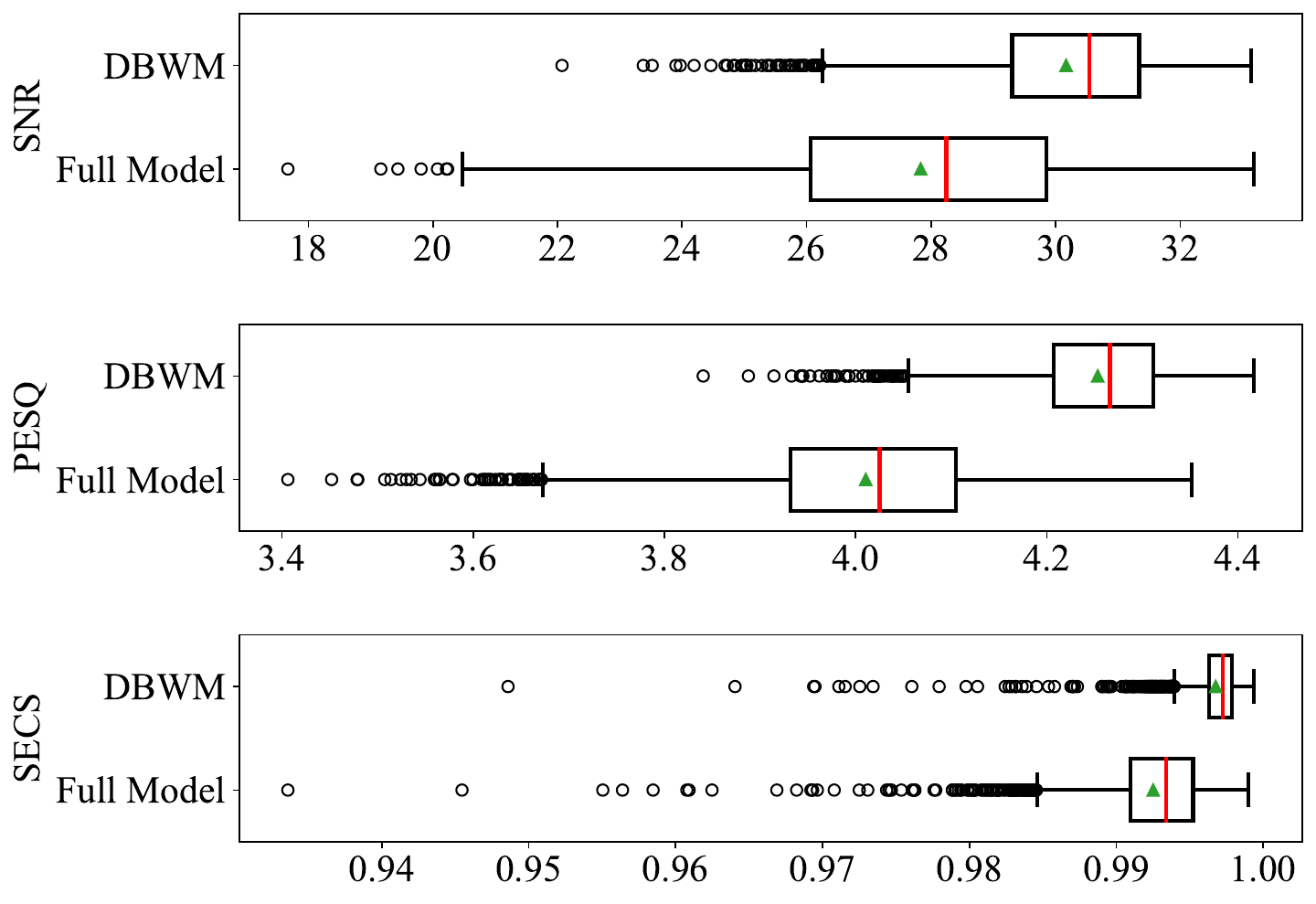}
    \vspace{-0.6em}
    \caption{Fidelity comparison between Full Model and the Distortion-Blind Watermarking Model (DBWM). Green triangles represent the mean values and red lines indicate the median values.}
    \vspace{-5pt}
    \label{fig:ab-fidelity}
\end{figure}

\para{Importance of Distortion Layer.}
To investigate the effectiveness of the modeled distortion layer, we remove it from our framework (as shown in \Fref{fig:framework}) and only jointly train watermark embedding and extraction to obtain a Distortion-Blind Watermarking Model (DBWM). 
As illustrated in \Fref{fig:ab-fidelity}, DBWM exhibits better fidelity
compared to the complete watermarking model (\ie, ``Full Model'').
\Tref{tb:ab} also indicates that the robustness of DBWM will significantly decrease in the scenarios of regular and low-quality voice cloning attacks.
This demonstrates that our framework without the distortion layer can achieve certain robustness against specific voice cloning attacks, and the appended distortion layer during training can further enhance the generalized robustness against more attacks. More results are displayed in  \ASref{sec:m_dbwm} with a consistent conclusion.
\hl{In \ASref{sec:appendix-c}, we also provide ablation studies on the skip concatenation and the masking ratio.}

\section{Discussion}
\label{sec:discussion}

\subsection{Comparison with Existing Audio Watermarking Methods} 
\label{sec:appendix-comparison}
To comprehensively showcase the efficacy of the proposed scheme, we conducted a comparative analysis with existing watermarking schemes. We use RFDLM \cite{liu2018patchwork}, FSVC \cite{zhao2021desynchronization}, and the proposed method to embed watermarks on the speech respectively, and then extract the watermarks from the synthesized speech based on the TTS model trained on this speech data respectively (VITS \cite{kim2021conditional} is utilized as the default voice cloning attack model). As shown in \Tref{tb:compare}, the existing audio watermarking schemes cannot resist the TTS model learning process and the watermarks cannot be retained in the synthesized speech, while the proposed scheme is better at retaining the watermark information.

\subsection{Embedding Different Watermarks for Verison Information} \label{sec:patrial}
In practice, the watermark may contain not only static information like names but also dynamic information like versions. In such a scenario, the collected audio by the attacker may have different watermarks, but we still want to extract the ownership information from the synthesized audio. To achieve it, we split the watermark bit string into two parts, where the former part includes static information and the latter part includes dynamic information. We watermark the original audio with two watermarks, whose former parts are the same while the latter parts are randomly different. Then, we adopt VITS for voice cloning attacks. As expected, we can extract the former static information with a 100\% ACC.
\hl{Similarly, we can easily replace version information with platform information to address the issue of multiple platforms.}
\hl{
In \ASref{sec:m_dis}, we provide more discussion on comparison with passive detection and the extension to the physical world.}

\begin{table}[]
\centering
\setlength\tabcolsep{8pt}
\caption{Comparison with traditional audio watermarking methods against voice cloning attacks. `syn' indicates metrics related to synthesized speech quality, while `wm' represents metrics related to watermark embedding and extraction.}
\vspace{-0.6em}
\begin{tabular}{ccccc}
\hline
Method & syn PESQ$\uparrow$   & syn SECS$\uparrow$   & wm SNR$\uparrow$    & wm ACC$\uparrow$           \\ \hline
FSVC \cite{zhao2021desynchronization}   & 0.9949     & 0.9139    & 21.1282 & 0.5554 ($\times$)    \\
RFDLM \cite{liu2018patchwork}  & 1.0303     & 0.9179    & 19.4668 & 0.5096 ($\times$)    \\
The Proposed   & 1.0342     & 0.9085     & 28.1650 & 1.0000 ($\checkmark$) \\ \hline
\end{tabular}
\label{tb:compare}
\vspace{-5pt}
\end{table}

%% file: 6_conclusion.tex
\section{Conclusion}
In the era of the Ear Economy, it is necessary to establish safeguards against the potential misuse of voice cloning technology. 
To this end, we propose a novel concept of ``Timbre Watermarking'',  based on which we design an end-to-end voice cloning detection framework. In this framework, we tailor
the watermark encoding with a repeated embedding strategy to obtain the inherent robustness against distortions in the time domain. In addition, we investigate different voice cloning attacks and find their shared process, \ie, normalization distortion, transformation distortion, and wave reconstruction distortion. Then, we incorporate them as a distortion layer into our framework to acquire generalization across different voice cloning attacks.
Extensive experiments demonstrate that the proposed ``Timbre Watermarking'' can achieve high robustness against common speech preprocessing distortions such as cropping, while also withstanding various voice cloning attacks and maintaining usability in real-world services such as PaddleSpeech, Voice-Cloning-App, and so-vits-svc. Moreover, we conduct ablation studies to explore the influence of watermark bits and the distortion layer. We expect our framework can shed some light on the future research of voice cloning detection and timbre protection.

\section*{Acknowledgement}
We thank anonymous reviewers for their constructive feedback. This work was supported in part by the Natural Science Foundation of China under Grant 62072421, 62002334, 62102386, 62121002, U20B2047 and Singapore Ministry of Education (MOE) AcRF Tier 2 MOE-T2EP20121-0006.

%% file: 7_appendix.tex
\appendix
\section{Appendix}

\subsection{More Discussion}
\label{sec:m_dis}
\subsubsection{Comparison with Passive Detection} 
\label{sec:det}
\begin{figure*}[t]
    \centering	
    \includegraphics[scale=0.4]{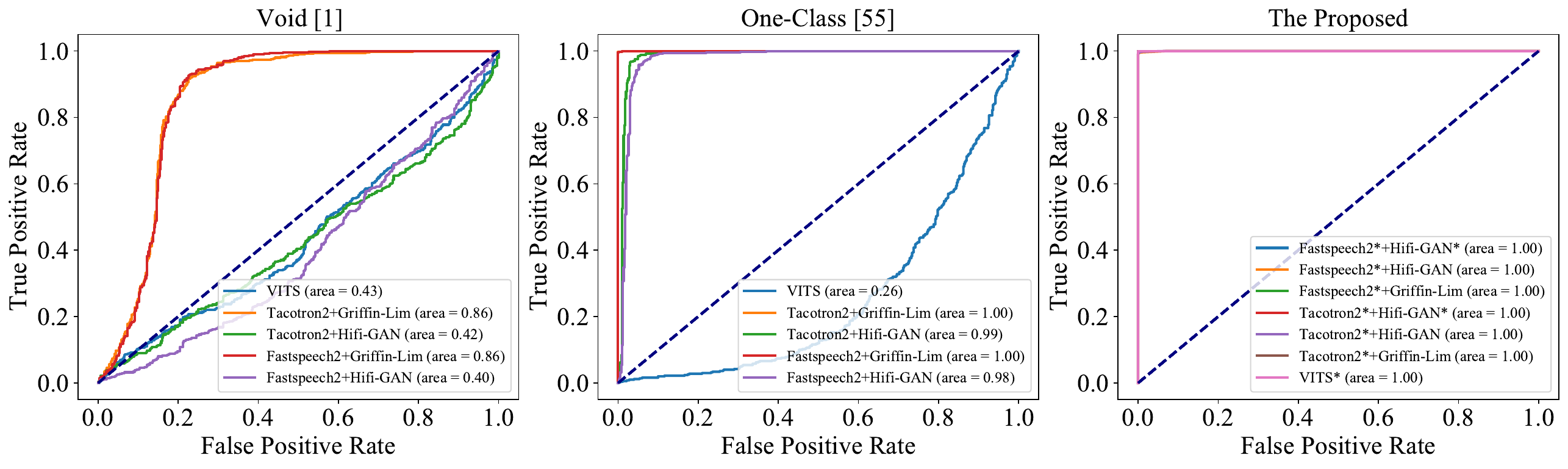}
    \caption{Comparison with state-of-the-art passive detection methods (Void \cite{ahmed2020void} and One-Class \cite{zhang2021one}). ROC curve against different voice cloning attacks are displayed.}
    \label{fig:roc}
        \vspace{-20pt}
\end{figure*}

\begin{figure}[htbp] 		
    \centering	
    \hspace*{18pt}
    \includegraphics[scale=0.7]{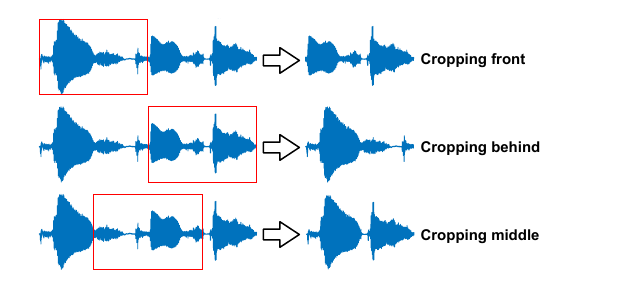}
    \vspace{-2mm}
    \caption{\hl{An example of a 50\% cropping with three strategies.}}
    \label{fig:crop-example}
\end{figure}

\begin{figure}[] 		
    \centering	
    \includegraphics[scale=0.35]{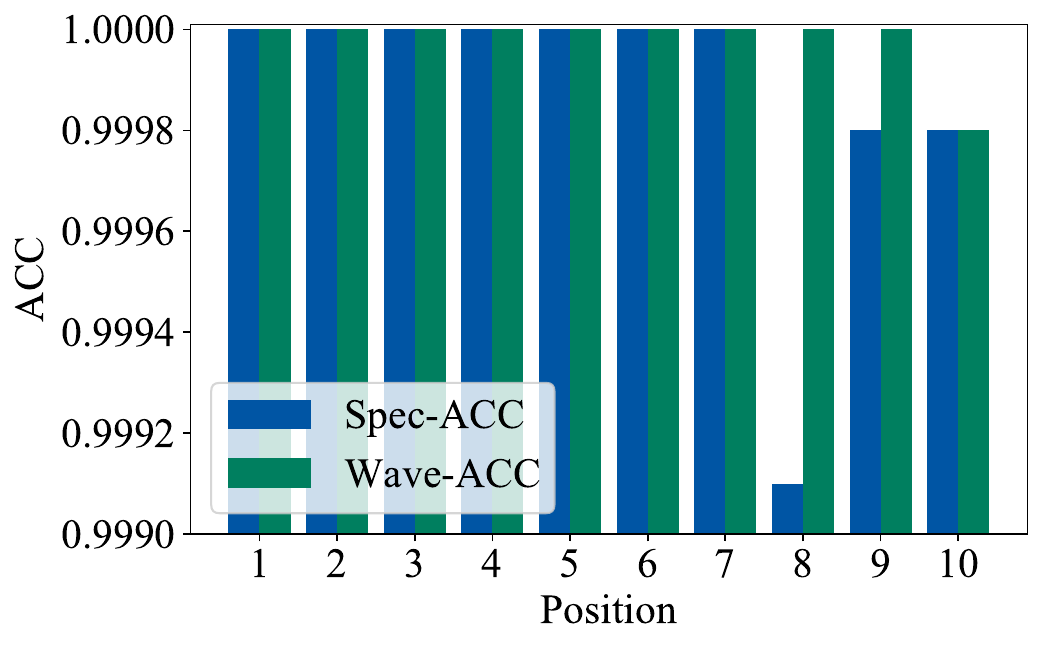}
    \vspace{-2mm}    
    \caption{ The watermark extraction accuracy of DBWM when different frequency bands of the spectrogram were masked. }
    \label{fig:spec-mask-DBWM}
                \vspace{-15pt}
\end{figure}

\begin{figure}[htbp] 		
    \centering	
    \includegraphics[scale=0.3]{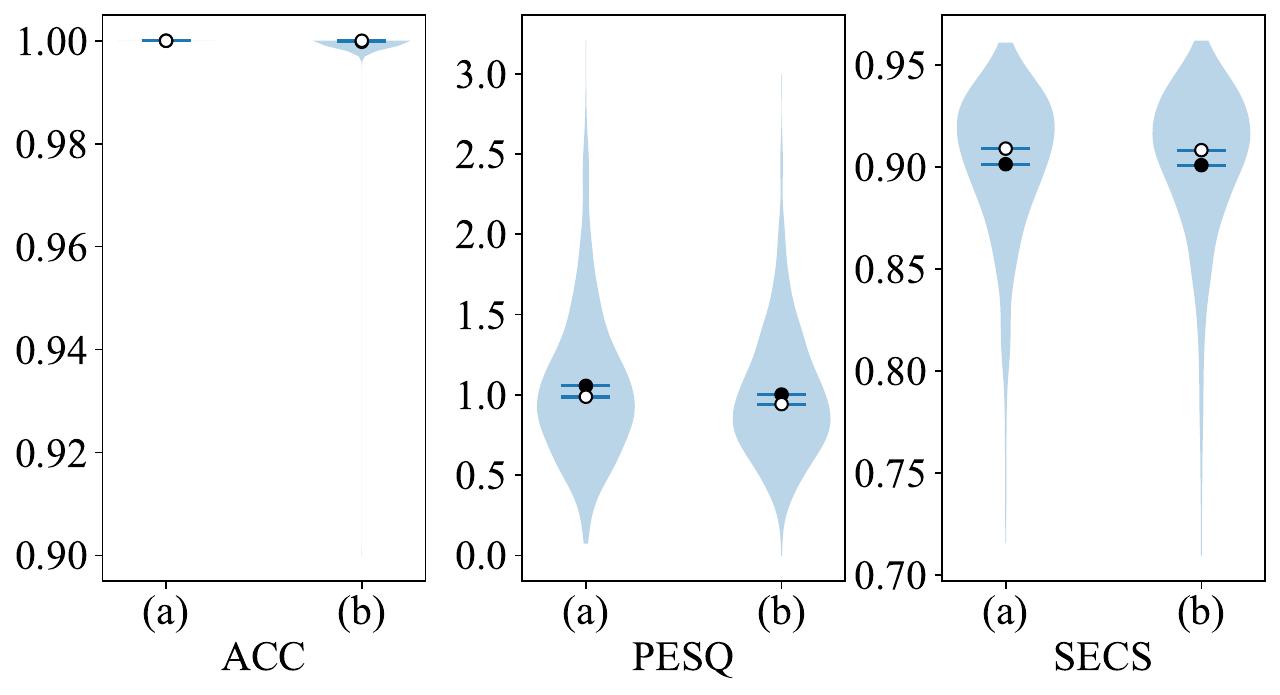}
    \vspace{-2mm}
    \caption{\hl{The performance of our method against voice cloning attacks with (a) / without (b) VAE reconstruction preprocessing.
    Black dots and white dots indicate mean value and median value, respectively.}}
    \label{fig:tango-vae}
\end{figure}

\begin{figure}[]
    \begin{center}
    \captionsetup[subfloat]{labelsep=none,format=plain,labelformat=empty}
    \begin{adjustbox}{margin=-3.5em -2em 0em 0em} 
    \subfloat{{\raggedleft\includegraphics[scale=0.25]{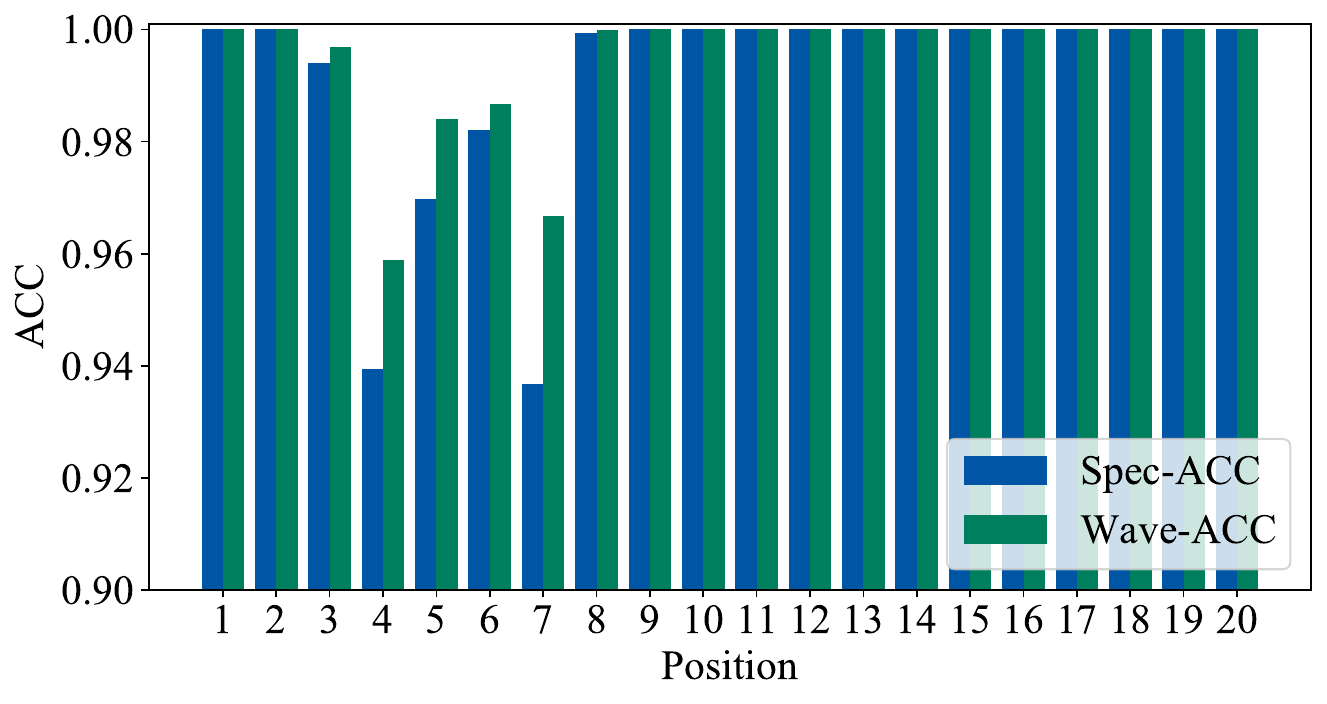}}}
    \end{adjustbox}
    \begin{adjustbox}{margin= 1em -2em -1em 0em}
    \subfloat{{\raggedright\includegraphics[scale=0.25]{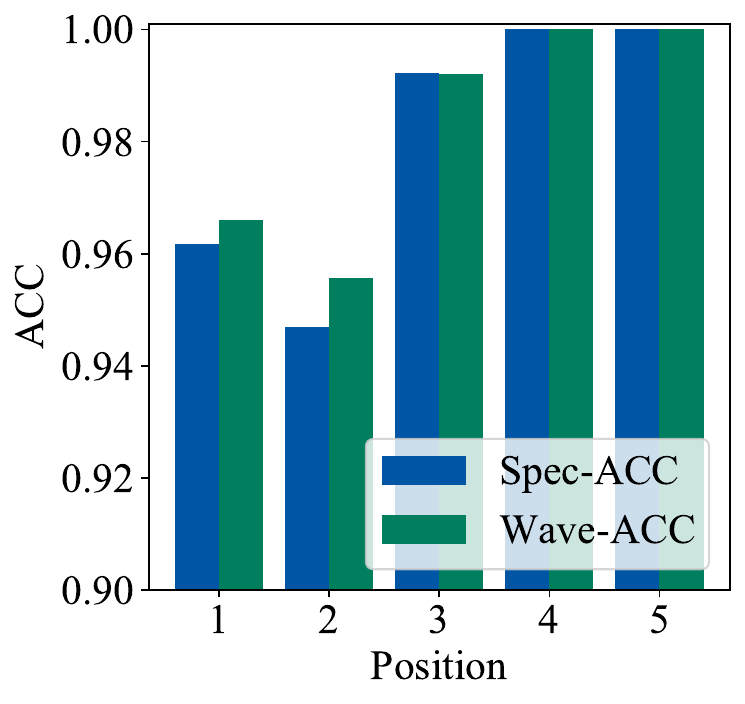}}}
    \end{adjustbox}\\ 
    \begin{adjustbox}{margin=-2em 0em 0em -1em}
    \subfloat[Mask 5\%]{{\raggedleft\includegraphics[scale=0.25]{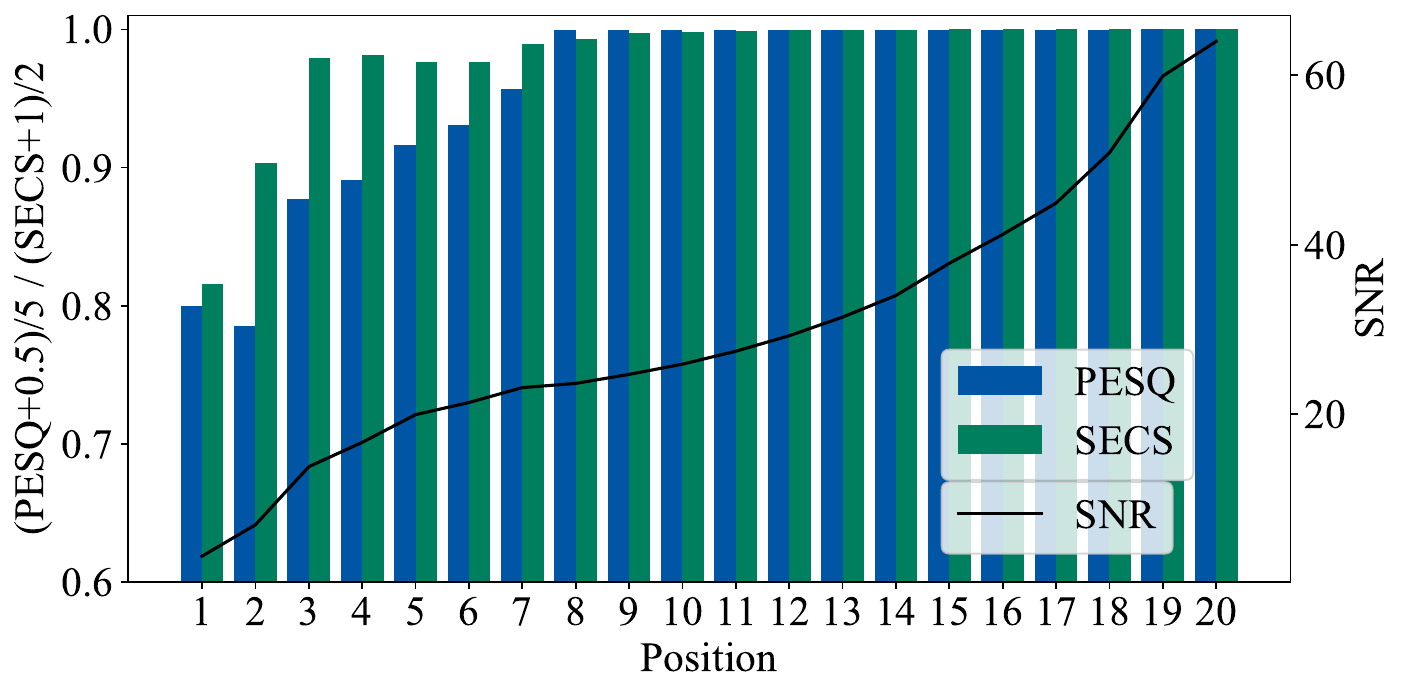}}}
    \end{adjustbox}
    \begin{adjustbox}{margin=0em 0em -1em -1em}
    \subfloat[Mask 20\%]{{\raggedright\includegraphics[scale=0.25]{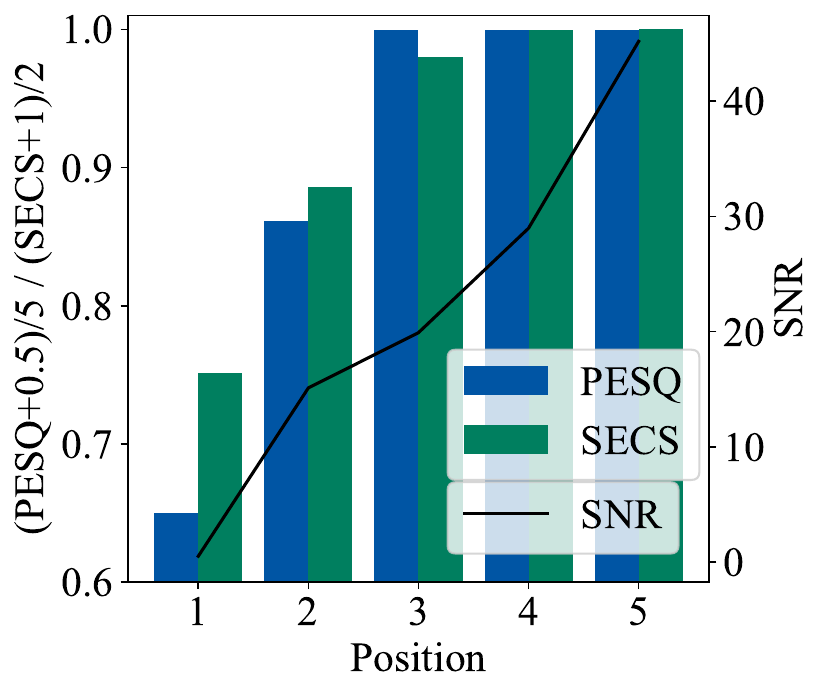}}}
    \end{adjustbox}
    \end{center}
    \caption{  \hl{Watermark extraction accuracy and speech quality under 5\% and 20\% masking ratios.}}
            \label{fig:mask}
\end{figure}

\begin{figure*}[t]
    \resizebox{\textwidth}{!}{
        \includegraphics[]{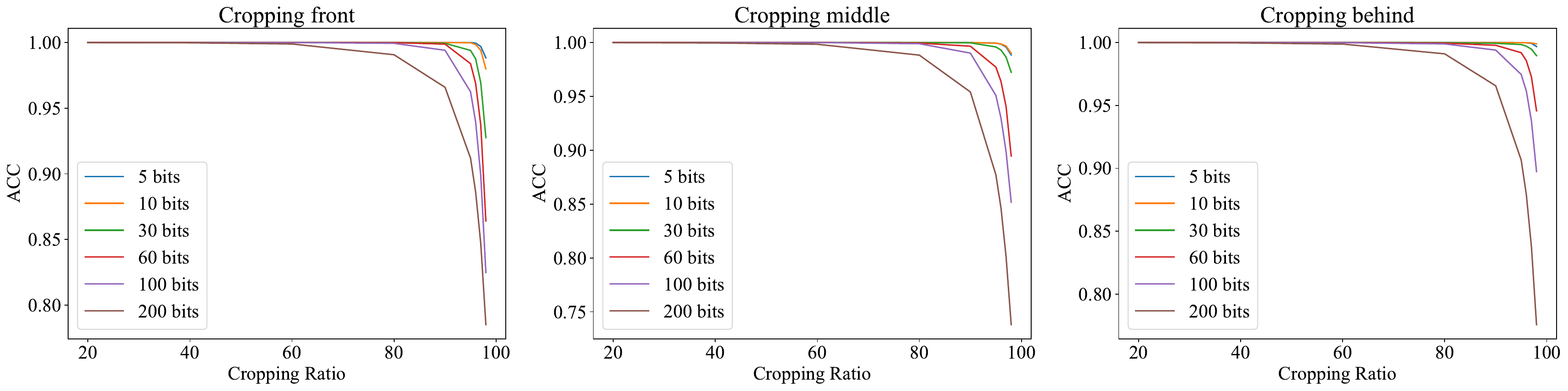}
    }
    \caption{\hl{Cropping robustness of the proposed method with different watermarking bits.}}
    \label{fig:crop-wm-bits}
\end{figure*}

\begin{figure}[]
\centering
    \begin{center}
        \captionsetup[subfloat]{labelsep=none,format=plain,labelformat=empty}
        \subfloat[Original]{
            {\raggedleft\includegraphics[scale=0.33]{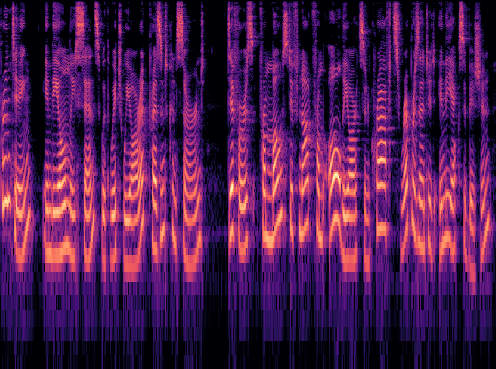}
            }
        }
        \subfloat[Real]{
            {\raggedleft\includegraphics[scale=0.33]{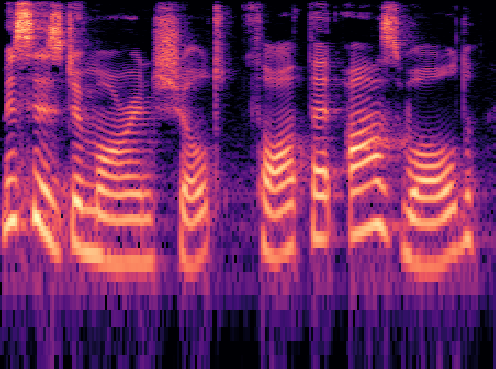}
            }
        }\\\vspace{-5pt}
        \subfloat[Watermarked]{
            {\raggedleft\includegraphics[scale=0.33]{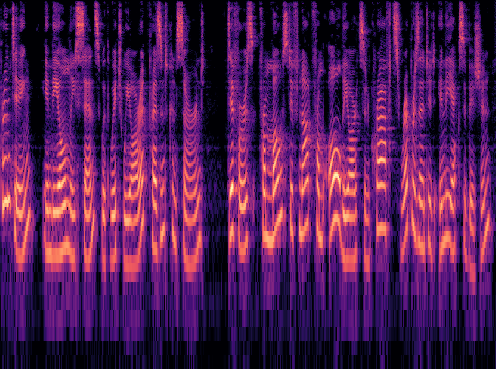}
            }
        }
        \subfloat[Cloned after watermarking]{
            {\raggedleft\includegraphics[scale=0.33]{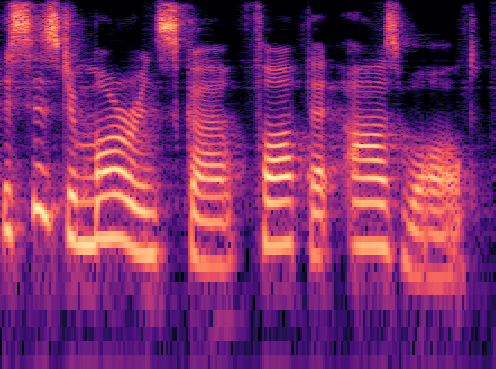}
            }
        }\\\vspace{-5pt}
        \subfloat[VAE reconstructed]{
            {\raggedleft\includegraphics[scale=0.33]{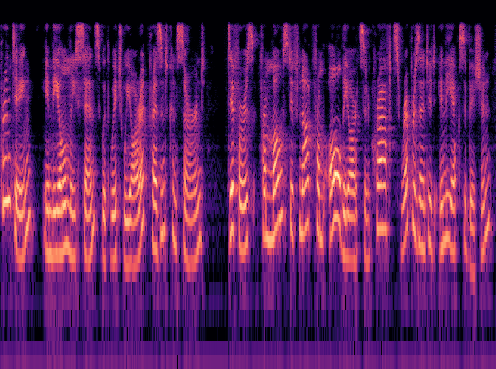}
            }
        }
        \subfloat[Cloned after VAE reconstruction]{
            {\raggedleft\includegraphics[scale=0.33]{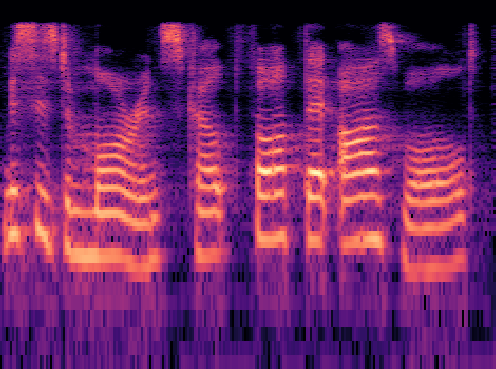}
            }
        }
    \end{center}
    \caption{\hl{The influence of VAE reconstruction on watermarked speech and cloned speech.}}
    \label{fig:tango-vae-spectrum}
\end{figure}

\begin{figure}[] 		
    \centering	
    \hspace{6pt}
    \includegraphics[scale=0.5]{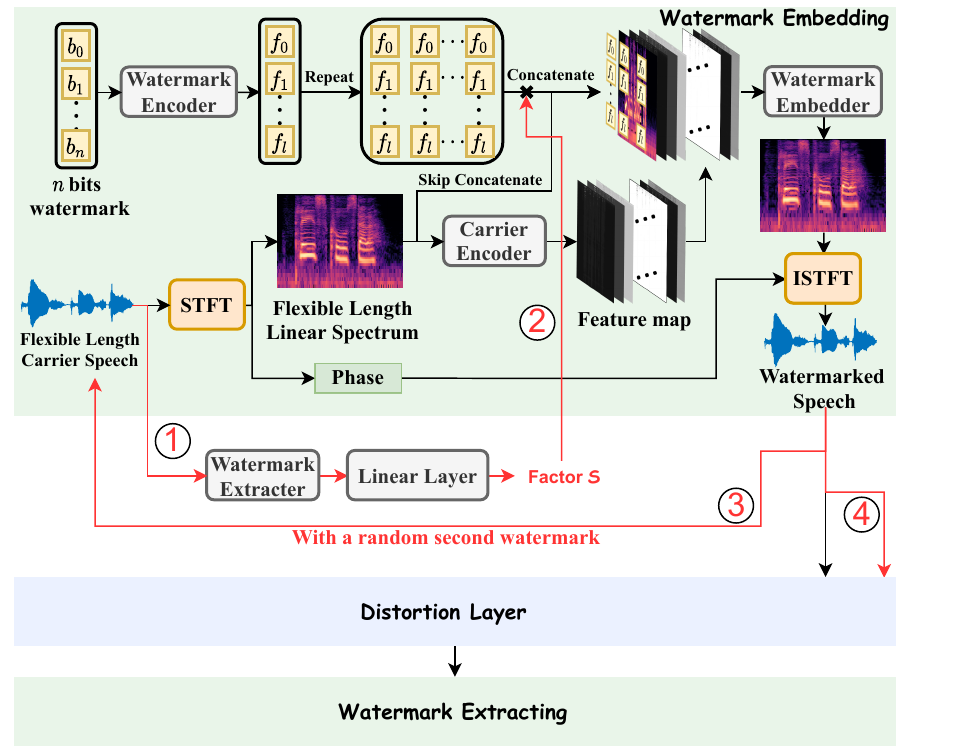}
    \caption{\hl{Overview of overwriting-resilient timbre watermark framework.}}
    \vspace{-15pt}
    \label{fig:framework-rewm}
\end{figure}

As mentioned above, passive detection usually cannot generalize to unseen synthesis methods. Two classic detection methods have been proposed to address the generalization limitation of data distribution and synthesis methods, namely, Void \cite{ahmed2020void} and One-Class \cite{zhang2021one}. We compare our method with these two detection approaches in terms of generalization against different voice cloning attacks. We follow the released unofficial code\footnote{\url{https://github.com/chislab/void-voice-liveness-detection}} and official code\footnote{\url{https://github.com/yzyouzhang/AIR-ASVspoof}} to reproduce these passive detection solutions.

Specifically, we measure the detection performance of these methods against different voice cloning attacks, where each attack generates 500 synthesized audios, respectively. For a fair comparison, we first divide the suspicious audio into 5 segments and set the accuracy threshold to 90\% for verification. We claim the detection is successful when all segments pass the verification. As shown in \Fref{fig:roc}, our method performs well in all cases, while two passive detection methods have poor performance ($AUC<0.5$) when facing synthesized audios generated by VITS \cite{kim2021conditional}.

\subsubsection{Extension to the Physical World}
{
In all the above experiments on voice cloning attacks, we consider that we can obtain suspicious synthesized speech in the digital world, even if the speech may be processed by some operations like MP3 compression. However, in some scenarios, we cannot directly obtain the digital version of the suspicious speech. Instead, we can only record the target speech for subsequent verification, which can be further distorted after the digital-physical-digital transformation. 
Fortunately, there are some works on audio watermarking and adversarial examples \cite{yuan2018commandersong, chen2020devil, liu2022dear}, which have demonstrated their effectiveness against such air-channel transmission distortion. A feasible way is to integrate them into the distortion layer to enhance robustness. We will explore this direction in future work.
}

\subsection{More Exploration on Robustness}
\label{sec:appendix-b}

\subsubsection{Robustness Against Cropping with Different Watermark Bits}
We further investigate the robustness of cropping processing to various watermark bit embedding quantities. As illustrated in \Fref{fig:crop-wm-bits}, there exists a trade-off between robustness and embedding capacity. Besides, an example of a 50\% cropping with different strategies are shown in \Fref{fig:crop-example}.

\begin{table}[htbp]
\centering
\caption{
The impact of different preprocessing on speech quality and the corresponding robustness of Distortion-Blind Watermarking Model (DBWM).}
\begin{tabular}{cccccc}
\hline
\multirow{2}{*}{Preprocessing}        & \multirow{2}{*}{Parameter}    & \multicolumn{3}{c}{Quality} & \multirow{2}{*}{ACC$\uparrow$} \\ \cline{3-5}
                                   &                           & SNR$\uparrow$      & PESQ$\uparrow$    & SECS$\uparrow$   &                      \\ \hline
\multirow{2}{*}{Resampling}          & 16   kHz               & 34.6674  & 4.4990  & 1.0000 & 0.9989               \\
                                   & 8 kHz                  & 16.6762  & 4.4986  & 0.9010 & 0.9169               \\ \hline
\multirow{4}{*}{Amplitude Scaling} & 20\%                   & 1.9382   & 4.4905  & 0.9590 & 1.0000               \\
                                   & 40\%                   & 4.4368   & 4.4968  & 0.9609 & 1.0000               \\
                                   & 60\%                   & 7.9589   & 4.4983  & 0.9778 & 1.0000               \\
                                   & 80\%                   & 13.9790  & 4.4989  & 0.9944 & 1.0000               \\ \hline
\multirow{8}{*}{MP3 Compression}   & 8   kbps               & 8.8171   & 2.1475  & 0.7591 & 0.6696               \\
                                   & 16 kbps                & 12.8881  & 3.3003  & 0.9566 & 0.9337               \\
                                   & 24 kbps                & 15.0170  & 3.8745  & 0.9890 & 0.9629               \\
                                   & 32 kbps                & 17.0359  & 4.0114  & 0.9960 & 0.9967               \\
                                   & 40 kbps                & 18.5867  & 4.1404  & 0.9975 & 0.9999               \\
                                   & 48 kbps                & 20.6921  & 4.2821  & 0.9986 & 1.0000               \\
                                   & 56 kbps                & 22.6389  & 4.3578  & 0.9990 & 1.0000               \\
                                   & 64 kbps                & 23.7704  & 4.3913  & 0.9992 & 1.0000               \\ \hline
Recount                            & 8   bps                & 22.8747  & 3.1435  & 0.9749 & 0.9041               \\ \hline
\multirow{4}{*}{Median Filtering}  & 5   Samples            & 14.3953  & 3.6013  & 0.9439 & 0.9714               \\
                                   & 15 Samples             & 8.6434   & 2.5087  & 0.7834 & 0.8750               \\
                                   & 25 Samples             & 5.2720   & 2.0886  & 0.7311 & 0.8183               \\
                                   & 35 Samples             & 3.1933   & 1.8187  & 0.6862 & 0.7701               \\ \hline
Low Pass Filtering                 & 2000 Hz                & 12.4479  & 3.8828  & 0.7252 & 0.7924               \\ \hline
High Pass Filtering                & 500  Hz                & 3.7917   & 3.8119  & 0.6578 & 1.0000               \\ \hline
\multirow{5}{*}{Gaussian Noise}    
                                   & 20 dB                  & 20.0001  & 3.0382  & 0.9090 & 0.7937               \\
                                   & 25 dB                  & 24.9994  & 3.4303  & 0.9667 & 0.8965               \\
                                   & 30 dB                  & 29.9977  & 3.7862  & 0.9920 & 0.9577               \\
                                   & 35 dB                  & 34.9934  & 4.0553  & 0.9981 & 0.9874               \\
                                   & 40 dB                  & 39.9871  & 4.2515  & 0.9993 & 0.9966               \\\hline
\end{tabular}
\vspace{-5pt}
\label{tb:common-distortion-DBWM}
\end{table}

\subsubsection{Robustness of DBWM}
\label{sec:m_dbwm}
We further investigate the robustness of DBWM in relation to common audio processing. As illustrated in \Tref{tb:common-distortion-DBWM}, without the distortion layer during training, the performance degrades in many cases.
In addition, we also investigate where watermarks are embedded by DBWM.  We conduct identical spectrogram masking experiments, as illustrated in \Fref{fig:spec-mask-DBWM}. The results reveal that DBWM predominantly embeds information within higher frequencies.

\hl{
\subsubsection{Countermeasurea Against Watermark Overwriting Attacks} \label{sec:appendix-overwriting}
As shown in \Tref{tb:rewm}, an adaptive attacker with access to the watermark encoder API can successfully conduct a self-overwriting attack using the original model. 
To address it, we design a mechanism to resist watermark overwriting of insider malicious users. We design a weighted embedding process, which is controlled by the subsequent watermark decoder, and add the watermark overwriting distortion in the original distortion layer to fine-tune the model. The final enhanced model can resist the re-watermark attack (ACC: 100\%).
The model structure is shown in \Fref{fig:framework-rewm}, which is composed of 4 steps: \circled{1} Extract the weight factor $S$ from the audio to be watermarked; \circled{2} Use factor $S$ to weight the watermark features in the watermark embedding network; \circled{3} Input the watermarked audio and a randomly generated second watermark into the watermark embedding network; \circled{4} Input the watermark overwritten audio into the distortion layer and proceed to subsequent watermark extraction and network parameter optimization.
We adopt VITS as the Voice cloning model. The watermark extraction accuracy (ACC) and the quality of all 500 synthesized speech are shown in \Tref{tb:exp-rewm}. Experimental results show that the overwriting-resilient strategy can resist watermark overwriting attacks even by internal attackers.
}

\begin{table}[t]
\begin{center}
\setlength\tabcolsep{7.5pt}
\caption{\hl{The comparison of robustness against watermark overwriting attacks between the default strategy and the overwriting-resilient strategy.} }
\begin{tabular}{ccccc}
\hline
Method       & PESQ$\uparrow$   & SECS$\uparrow$    & wm1-ACC$\uparrow$       & wm2-ACC$\downarrow$         \\ \hline
Default      & 0.9891           & 0.8968            & 0.4000 ($\times$)       & 1.0000 ($\checkmark$)       \\
Overwriting-resilient     & 1.0269           & 0.9145            & 1.0000 ($\checkmark$)   & 0.4000 ($\times$)           \\
\hline
\end{tabular}
\vspace{-25pt}
\label{tb:exp-rewm}
\end{center}
\end{table}

\subsection{More Ablation Studies}
\label{sec:appendix-c}

\hl{
\para{The Influence of Different Masking Ratios.} \label{sec:mask-ratio-2}
See \Fref{fig:mask}.}

\subsection{Details of Network Architectures} \label{sec:structure}
\Tref{tb:structure-ow} shows the structure of the attacker's watermark overwriting model, where ReluBlock contains a convolutional layer, an InstanceNorm, and the LeakyReLU activation function.

\begin{table}[htbp]
\centering
\caption{The detailed structure of each sub-network of the attacker's model for watermark overwriting, where ``wm\_length" represents the length of the watermark bits.}
\setlength\tabcolsep{2pt}
\begin{tabular}{cccc}
\hline
                                     & Groups             & Input   channel/dim & Output   channel/dim \\ \hline
Watermark   Encoder                  & Linear + LeakyReLU & wm\_length          & 513                  \\ \hline
\multirow{6}{*}{Carrier Encoder}     & ReluBlock          & 1                   & 64                   \\
                                     & ReluBlock          & 64                  & 64                   \\
                                     & ReluBlock          & 64                  & 64                   \\
                                     & ReluBlock          & 64                  & 64                   \\
                                     & ReluBlock          & 64                  & 64                   \\
                                     & ReluBlock          & 64                  & 64                   \\ \hline
\multirow{4}{*}{Watermark Embedder}  & ReluBlock          & 66                  & 64                   \\
                                     & ReluBlock          & 64                  & 64                   \\
                                     & ReluBlock          & 64                  & 64                   \\
                                     & ReluBlock          & 64                  & 1                    \\ \hline
\multirow{5}{*}{Watermark Extracter} & ReluBlock          & 1                   & 64                   \\
                                     & ReluBlock          & 64                  & 64                   \\
                                     & ReluBlock          & 64                  & 64                   \\
                                     & ReluBlock          & 64                  & 64                   \\
                                     & ReluBlock          & 64                  & 64                   \\
                                     & ReluBlock          & 64                  & 1                    \\ \hline
Watermark   Decoder                  & Linear             & 513                 & 1                    \\ \hline
\multirow{4}{*}{\hl{Discriminator}}       & \hl{ReluBlock}          & \hl{1}                   & \hl{16}              \\
                                     & \hl{ReluBlock}          & \hl{16}                  & \hl{32}              \\
                                     & \hl{ReluBlock}          & \hl{32}                 & \hl{64}             \\
                                     & \hl{Linear}             & \hl{64}               & \hl{1}           \\ \hline
\end{tabular}
\label{tb:structure-ow}
\end{table}

\subsection{Phrases for Synthesis}
\Tref{tb:phrases} and \Tref{tb:phrases-cns} list the English and Chinese phrases, respectively, used for synthetic speech generation in real-life scenarios using voice cloning tools.

\begin{table}[htbp]
\centering
\caption{English phrases used for real-world services in \Sref{sec:vc-tools}.}
\tiny
\begin{tabular}{c >{\centering\arraybackslash}m{7.5cm}}
\hline
1. & There is, according to legend, a boiling pot of gold at one end.\\ \hline
2. & People look, but no one ever finds it.\\ \hline
3. & Throughout the centuries people have explained the rainbow in various ways.\\ \hline
4. & Some have accepted it as a miracle without physical explanation.\\ \hline
5. & To the Hebrews it was a token that there would be no more universal floods.\\ \hline
6. & Since then physicists have found that it is not reflection, but refraction by the raindrops which causes the rainbows.\\ \hline
7. & Many complicated ideas about the rainbow have been formed.\\ \hline
8. & The difference in the rainbow depends considerably upon the size of the drops, and the width of the colored band increases as the size of the drops increases.\\ \hline
9. & If the red of the second bow falls upon the green of the first, the result is to give a bow with an abnormally wide yellow band, since red and green light when mixed form yellow.\\ \hline
10. & This is a very common type of bow, one showing mainly red and yellow, with little or no green or blue.\\ \hline
\end{tabular}
\label{tb:phrases}
\end{table}

\begin{CJK}{UTF8}{gbsn}
\begin{table}[htbp]
\begin{center}
\caption{Chinese phrases used for PaddleSpeech \cite{PaddleSpeech}   in \Sref{sec:vc-tools}.}
\tiny
\begin{tabular}{c >{\centering\arraybackslash}m{7.5cm}}
\hline
1. & 正 因 为此 斯诺 曾 感触 颇 深 地说 鲁迅 虽 身材 瘦弱 矮小 但 和 鲁迅 在一起 你 必须 仰视 着 去 领会 那 崇高 的 思想 \\\hline 
2. & 韩 电 位于 西北 电网 末端 其 安全生产 对 该 电网 稳定 运行 有 重要意义                              \\ \hline
3. & 如果 下到 谷底 顿觉 寒气 袭人 抬头 四 顾 阴森森 的 土 壁 确有 群 魔 压 顶 的 惊心动魄 之 感                \\ \hline
4. & 早稻 播种 和 育秧 的 天气 条件 有利 与否 与 这 一 期间 的 日 平均 温度 阴雨 日数 密切 相关                 \\\hline
5. & 六年 来 神池 道情 剧团 台上 为 矿工 认真 演出 台下 义务 慰问 老 矿工                               \\ \hline
6. & 一群 如 火球 的 娃娃 便 滚动 进场 每人 持 两 朵 大 黄花 飞舞 似 风吹 花圃 静 立 若 菊园 吐 香              \\ \hline
7. & 就 在 那 阴沟 旁边 却 高 高下 下放 着 几 盆花 也有 夹竹桃 也有 常青 的 盆栽                          \\ \hline
8. & 昔日 的 高 母 村 七百 口 人 仅 光棍 汉 就 有 二百多 庄 户 人 穷 得 连 个 称 盐 打油 的 钱 也 没有          \\ \hline
9. & 发射 约 二 十分钟 后 飞行 样机 打开 降落伞 溅 落在 北纬 二十 八度 五 十一分 东经 一百四十 三度 四十分 的 太平洋 海 面上 \\ \hline
10. & 杭城 某 丝织 厂 一 位 沈 姓 姑娘 怀揣 一千 多元 钱 来到 濮 院 选购 羊毛衫                           \\\hline
\end{tabular}
\label{tb:phrases-cns}
\end{center}
\end{table}
\end{CJK}